\documentclass[aps,preprint,superscriptaddress,prb]{revtex4-2}
\usepackage{amsmath}
\usepackage{amssymb}
\usepackage{graphicx}
\usepackage[english]{babel}
\usepackage{bm}
\usepackage{mathrsfs}

\newcommand{\op}[2][]{\hat{b}_{{\mathbf{#2}}_{#1}}}
\newcommand{\opc}[2][]{\hat{b}^{\dagger}_{{\mathbf{#2}}_{#1}}}

\begin{document}

\title{Beyond the Fr\"ohlich Hamiltonian: Path integral treatment of large polarons in anharmonic solids}

\author{Matthew Houtput}
\affiliation{Theory of Quantum Systems and Complex Systems, Universiteit Antwerpen, B-2000 Antwerpen, Belgium}

\author{Jacques Tempere}
\affiliation{Theory of Quantum Systems and Complex Systems, Universiteit Antwerpen, B-2000 Antwerpen, Belgium}

\begin{abstract}

The properties of an electron in a typical solid are modified by the interaction with the crystal ions, leading to the formation of a quasiparticle: the polaron. Such polarons are often described using the Fr\"ohlich Hamiltonian, which assumes the underlying lattice phonons to be harmonic. However, this approximation is invalid in several interesting materials, including the recently discovered high-pressure hydrides which superconduct at temperatures above $200$K. In this paper, we show that Fr\"ohlich theory can be extended to eliminate this problem.

We derive four additional terms in the Fr\"ohlich Hamiltonian to account for anharmonicity up to third order. We calculate the energy and effective mass of the new polaron, using both perturbation theory and Feynman's path integral formalism. It is shown that the anharmonic terms lead to significant additional trapping of the electron. The derived Hamiltonian is well-suited for analytical calculations, due to its simplicity and since the number of model parameters is low. Since it is a direct extension of the Fr\"ohlich Hamiltonian, it can readily be used to investigate the effect of anharmonicity on other polaron properties, such as the optical conductivity and the formation of bipolarons.

\end{abstract}

\maketitle

\section{Introduction}
The polaron concept was introduced by Landau \cite{Landau1933} and Pekar \cite{Landau1948}
to explain the dynamics of an electron interacting with a crystal lattice. Classically,
the electron-phonon interaction can be explained by the electron displacing the ions out of their
equilibrium positions, creating a polarization field that interacts with the electron.
The entire system can be described as a quasiparticle: the polaron \cite{Landau1933}.
It is one of the simplest models describing an impurity interacting with a boson bath, and thus
finds applications in many other fields of physics. Specific examples include spin polarons
\cite{Nagaev1974}, magnetic polarons \cite{Koepsell2019}, exciton polarons
\cite{Verzelen2002}, and the Bose \cite{Jorgensen2016,Shchadilova2016,Ichmoukhamedov2019}
and Fermi \cite{Schirotzek2009} polaron in ultracold gases.

Usually, the harmonic approximation is made when discussing polarons.
One can assume that the lattice potential around an ion's equilibrium position is
approximately quadratic, so the restoring forces are linear.
This results in a bath of phonons that do not interact directly with
each other.
This approximation is justified in most materials as the phonon amplitude
is usually small.
If the electron wavefunction extends over many unit cells, the lattice can be viewed
as a continuous field, and the polaron is called `large'. 
The Hamiltonian for 
`large' polarons in the harmonic approximation is known as the Fr\"ohlich Hamiltonian; it is one of the
simplest non-trivial Hamiltonians of quantum field theory.
Treating the electron in first quantization and the phonons in second quantization,
it reads \cite{Frohlich1954}:
\begin{align}
    \hat{H} & = \hat{H}_{\text{e}} + \hat{H}_{\text{ph}} + \hat{H}_{\text{e}-\text{ph}}, \nonumber \\
    & =  \frac{\hat{\mathbf{p}}_{\text{el}}^2}{2m}+ \sum_{\mathbf{k}} \hbar \omega_{\mathbf{k}} \left( \opc{k} \op{k} + \frac{1}{2} \right) + \sum_{\mathbf{k}} \left(V_{\mathbf{k}} \opc{k} e^{-i \mathbf{k}\cdot\hat{\mathbf{r}}_{\text{el}}} + V_{\mathbf{k}}^* \op{k} e^{i \mathbf{k}\cdot \hat{\mathbf{r}}_{\text{el}}} \right).  \label{FrohHam}
\end{align}
Here $\omega_{\mathbf{k}}$ is the phonon dispersion and $V_{\mathbf{k}}$
is the electron-phonon interaction strength: both are functions of the
phonon momentum $\mathbf{k}$. The specific form of these functions depends
on the system at hand and can significantly alter the underlying
physics of the problem \cite{Grusdt2015,Grusdt2017}. The operators
$\opc{k}$ and $\op{k}$ create and annihilate a phonon with wavenumber ${\mathbf{k}}$,
respectively.
A defining characteristic of the Fr\"ohlich
Hamiltonian is that $\hat{H}_{\text{ph}}$ is quadratic and
$\hat{H}_{\text{e}-\text{ph}}$ is linear in the phonon operators.

In reality, the lattice potential is not harmonic, which must be considered
when looking at high-pressure hydrides \cite{Errea2013,Errea2014,Errea2015}.
In the classical picture, since the
mass of hydrogen ions is small, the phonon amplitude will be too large for
the harmonic approximation to apply. Interest in high-pressure hydrides has
been strongly renewed since the discovery of high-temperature superconductivity
in sulphur hydride \cite{Drozdov2015} ($T_c = 203$K), lanthanum hydride
\cite{Somayazulu2019} ($T_c = 260$K), and carbonaceous sulphur hydride
\cite{Snider2020} ($T_c = 288$K) when these materials are put under megabar
pressures. Similarly, pure hydrogen has been theoretically predicted to metallize and
superconduct at room temperature under the high pressure \cite{Ashcroft1968, Dias2017, Loubeyre2019}.
Superconductivity in these materials appears to be conventional and thus phonon-mediated \cite{Drozdov2015}.
However, the harmonic approximation is not applicable
\cite{Errea2013,Errea2014,Errea2015}, so additional ``anharmonic'' terms must be
considered in the electron-phonon Hamiltonian \eqref{FrohHam}.

Most of the research on anharmonic polarons focuses on `small' polarons
\cite{Zolotaryuk1998, Voulgarakis2000, Velarde2010}, where the electron is localized
around a single lattice atom. The most recent,
and to our knowledge only, investigation of the anharmonic terms for
large polarons is due to Kussow \cite{Kussow2009}. In \cite{Kussow2009}, the dominant
anharmonic term for the Fr\"ohlich Hamiltonian \eqref{FrohHam} is derived, and the polaron
energy is calculated using perturbation theory in the weak coupling regime.
However, the Hamiltonian is only useful for qualitative calculations due to several
assumptions and errors in its derivation. In this paper, we redo the
derivation presented in \cite{Kussow2009}, fixing these errors and including the 3-phonon terms.
Additionally, we will calculate the polaron energy using Feynman's
path integral method \cite{Feynman1955}, allowing us to look at the intermediate coupling
and strong coupling regimes as well. The presented Hamiltonian can be used to calculate
polaron properties in high-pressure hydrides \cite{Drozdov2015,Somayazulu2019,Snider2020}, but also in anharmonic semiconductors such as boron nitride \cite{Brito2019}
and aluminium nitride \cite{Shulumba2016,Yaddanapudi2018}.

The structure of this paper is as follows. We derive additional anharmonic terms
in the Fr\"ohlich hamiltonian \eqref{FrohHam} in section \ref{Sec:Hamiltonian}.
In sections \ref{Sec:Perturbation}-\ref{Sec:Feynman}
the ground state energy and effective mass of a single large anharmonic polaron
are calculated, using perturbation theory in section \ref{Sec:Perturbation} and
Feynman's path integral method \cite{Feynman1955} in section \ref{Sec:Feynman}.
We summarize our findings in section \ref{Sec:Conclusions}.

\section{The anharmonic polaron Hamiltonian} \label{Sec:Hamiltonian}
\subsection{Derivation}\label{Sec:Derivation}
Here, we rederive the Hamiltonian based on the derivations of Fr\"ohlich \cite{Frohlich1954}
and Kussow \cite{Kussow2009}. We assume the same model system used in both of these derivations:
an ionic, polarizable lattice with two ions in the primitive unit cell.
The masses of the two ions are denoted with $m_1$ and $m_2$. One or more electrons with band mass
$m$ and charge $-e$ are placed in this lattice at positions $\mathbf{r}_{\text{el},i}$.
We assume that electron-phonon coupling is dominated by the longitudinal optical
(LO) phonons, so all other phonon contributions are neglected. The displacements
of the ions from their equilibrium positions are denoted by $\mathbf{r}_1$ and $\mathbf{r}_2$,
where the position of each ion is measured relative to its respective equilibrium
position. The kinetic energy per unit volume due to these displacements is given by:
\begin{equation}
    E_{k} = \frac{1}{V_0} \left( \frac{1}{2} m_1 \Dot{\mathbf{r}}_1^2 + \frac{1}{2} m_2 \Dot{\mathbf{r}}_2^2 \right),
\end{equation}
where $V_0$ is the volume of the unit cell. We now switch to center-of-mass coordinates.
The movement of the center of mass leads to acoustic phonons and can therefore be neglected \cite{Kussow2009}.
The kinetic energy density can then be written in terms of only the relative
displacement $\mathbf{w}$:
\begin{align}
    E_k & = \frac{1}{2} \Dot{\mathbf{w}}^2, \label{Ekin} \\
    \mathbf{w} & := \sqrt{\frac{1}{V_0} \frac{m_1 m_2}{m_1 + m_2}} (\mathbf{r_2}-\mathbf{r_1}).
\end{align}
Since we consider large polarons, the lattice can be approximated by a polarizable
continuum. Mathematically, this means we can treat $\mathbf{w} = \mathbf{w}(\mathbf{r})$
as a position-dependent vector field. 

Aside from the kinetic energy, the lattice also has an interaction energy $U$
per unit volume. This internal energy contains the interaction energy of the ions,
but also a contribution due to the electric displacement field $\mathbf{D}$ which
is solely due to the electrons. The internal energy is a function of $\mathbf{w}$
and $\mathbf{D}$ \cite{Gurevich1986}, and satisfies:
\begin{equation} \label{dU}
    dU = -\ddot{\mathbf{w}} \cdot d\mathbf{w} + \mathbf{E} \cdot d\mathbf{D}.
\end{equation}
Here $\ddot{\mathbf{w}}$ is proportional to the force on the atoms, and $\mathbf{E}$
is the electric field. Both $\mathbf{D} = \mathbf{D}(\mathbf{r},t)$ and
$\mathbf{E} = \mathbf{E}(\mathbf{r},t)$ are position- and time-dependent, just like
the relative displacement $\mathbf{w}(\mathbf{r},t)$; however, from now on, we will
drop this explicit dependence and simply write $\mathbf{D}$, $\mathbf{E}$, and $\mathbf{w}$.

Since $\mathbf{w} = \mathbf{D} = \mathbf{0}$ corresponds to the equilibrium position
of the lattice, the function $\tilde{U}$ can be expanded in powers of $\mathbf{w}$ and
$\mathbf{D}$. The first non-trivial order is an expansion up to second order, which is
the harmonic expansion that will yield the Fr\"ohlich Hamiltonian \eqref{FrohHam}.
In this paper, we consider the internal energy up to third order. It can be written as:
\begin{align}
    U(\mathbf{w},\mathbf{D}) & \approx \frac{1}{2} \gamma^{(0)}_{ij} w_i w_j + \gamma^{(1)}_{ij} w_i D_j + \frac{1}{2} \gamma^{(2)}_{ij} D_i D_j \nonumber \\
    & + \frac{1}{6} A^{(0)}_{ijl} w_i w_j w_l + \frac{1}{2} A^{(1)}_{ijl} w_i w_j D_l + \frac{1}{2} A^{(2)}_{ijl} w_i D_j D_l + \frac{1}{6} A^{(3)}_{ijl} D_i D_j D_l.  \label{UInternal}
\end{align}
where the indices $i,j,l$ can take the values in $\{x,y,z\}$. We use the Einstein
summation convention of implied summation over repeated indices throughout the remainder of
this article.
Contrary to Kussow \cite{Kussow2009}, we expand the internal energy as a function
of $\mathbf{D}$ instead of $\mathbf{E}$. The two methods are equivalent since the expansion
coefficients are related to each other. The final term is proportional to $\mathbf{D}^3$ and is
responsible for the nonlinear optical response of the material. It can be neglected in most materials,
but it will be carried here for completeness.

In this expression, second order tensors $\gamma^{(n)}_{ij}$ and third order tensors
$A^{(n)}_{ijk}$ appear.  These tensors can be interpreted as material parameters:
in fact, in section \ref{Sec:Cubic} we will show that $\gamma_{ij}^{(0)}$, $\gamma_{ij}^{(1)}$
and $\gamma_{ij}^{(2)}$ can be written in terms of measurable quantities for a cubic crystal.
In general, these parameters can be calculated using \textit{ab initio} methods by
calculating the mixed partial derivatives of the internal energy with respect to $\mathbf{w}$
and $\mathbf{D}$. The tensors $\gamma^{(0)}_{ij}$, $\gamma^{(2)}_{ij}$, $A^{(0)}_{ijk}$
and $A^{(3)}_{ijk}$ are totally symmetric, and the tensors $A^{(1)}_{ijk}$ and $A^{(2)}_{ijk}$
are symmetric in one pair of their indices:
\begin{align}
    A^{(1)}_{ijk} = & A^{(1)}_{jik}, \label{A1symmetry} \\
    A^{(2)}_{ijk} = & A^{(2)}_{ikj}. \label{A2symmetry}
\end{align}
In the most general case, the tensor $\gamma^{(1)}_{ij}$ has no symmetry.

We now introduce $N$ free electrons in the system. We assume a parabolic energy dispersion with
band mass $m$. Their kinetic energy takes the standard form:
\begin{equation} \label{ElectronEnergy}
E_{\text{el}} = \sum_{i=1}^{N}\frac{\mathbf{p}^2_{\text{el},i}}{2m},
\end{equation}
We note that this form is only valid for cubic crystals, but the band mass $m$ can readily
be replaced with an effective inverse mass tensor to account for anisotropy.
Integrating the energy densities \eqref{Ekin}, \eqref{UInternal} over the crystal volume
$V$ and adding the electron kinetic energy \eqref{ElectronEnergy}, we obtain the classical
Hamiltonian of the system:
\begin{equation} \label{HamClassical}
H = \sum_{i=1}^{N}\frac{\mathbf{p}^2_{\text{el},i}}{2m} + \int_V \frac{1}{2} \Dot{\mathbf{w}} \cdot \Dot{\mathbf{w}} d^3 \mathbf{r} + \int_V U(\mathbf{w},\mathbf{D}) d^3\mathbf{r}.
\end{equation}
All that remains is to find expressions for the phonon field $\mathbf{w}$ and the electric
displacement field $\mathbf{D}$.

The longitudinal component of the electric displacement field is only due to the electrons.
Its transverse component is zero \cite{Frohlich1954}, since $\mathbf{E}$ and the polarization
field $\mathbf{P}$ are both longitudinal: this follows from the quasi-static third Maxwell
equation and the fact that we consider longitudinal phonons, respectively.
Therefore, the electric displacement field has an analytical expression \cite{Frohlich1954}:
\begin{equation} \label{Ddef}
    \mathbf{D}(\mathbf{r}) = \sum_{i=1}^N \frac{e}{4 \pi} \bm{\nabla} \left( \frac{1}{|\mathbf{r}-\mathbf{r}_{\text{el},i}|} \right) .
\end{equation}
For future calculations, it will be useful to write the displacement field in Fourier space as follows:
\begin{equation} \label{Dfourier}
\mathbf{D}(\mathbf{r}) = -\frac{i e}{V} \sum_{\mathbf{k}\neq\mathbf{0} }  \frac{\mathbf{n}^{\mathbf{k}}}{|\mathbf{k}|} \rho_{\mathbf{k}}  e^{-i \mathbf{k}\cdot\mathbf{r}},
\end{equation}
where
\begin{equation}
\rho_{\mathbf{k}} = \sum_{i=1}^N e^{i \mathbf{k} \cdot \mathbf{r}_{\text{el},i}}
\end{equation}
is the density operator of the electrons, and we also introduced the symbol
$\mathbf{n}^{\mathbf{k}} = \frac{\mathbf{k}}{|\mathbf{k}|}$ for the unit vector in the direction of $\mathbf{k}$. We will write their components as:
\begin{equation}
n_i^{\mathbf{k}} = \frac{k_i}{|\mathbf{k}|}.
\end{equation}
These unit vectors will feature often in our calculations and results.

Up to a proportionality constant, the field $\mathbf{w}$ can be interpreted as a phonon coordinate.
Its conjugate momentum is simply $\frac{\partial H}{\partial \dot{\mathbf{w}}} = \dot{\mathbf{w}}$.
We can quantize $\mathbf{w}$ and $\dot{\mathbf{w}}$ using the ladder operators $\op{k}$
and $\opc{k}$, if we can identify the phonon frequencies. Rather than to derive an equation
for the polarization density $\mathbf{P}$ as is done in \cite{Frohlich1954} and \cite{Kussow2009},
we do this by looking at the Hamiltonian \eqref{HamClassical} in the case of no electrons:
$\mathbf{D} = \mathbf{0}$.
Furthermore, we only look at the harmonic approximation, so we only use the first
line of equation (\ref{UInternal}). The Hamiltonian density then takes the form of a
harmonic oscillator:
\begin{equation}
\mathcal{H} = \frac{1}{2} \Dot{\mathbf{w}} \cdot \Dot{\mathbf{w}} + \frac{1}{2} \mathbf{w} \cdot \bm{\gamma}_0 \cdot \mathbf{w},
\end{equation}
where the bold $\bm{\gamma}_0$ indicates the matrix with components $\gamma^{(0)}_{ij}$.
Since the matrix $\bm{\gamma}_0$ is symmetric, it can be diagonalized. Its eigenvectors are
the eigendirections of the phonons, and its
eigenvalues $\omega_i^2$ are the squares of the phonon frequencies. Therefore, if we write
$\bm{\gamma}_0 = \bm{\Omega}^2$, the matrix $\bm{\Omega}$ can be used instead of the phonon
frequency in the definition of the ladder operators.

\begin{figure}
\centering
\includegraphics[width=8.6cm]{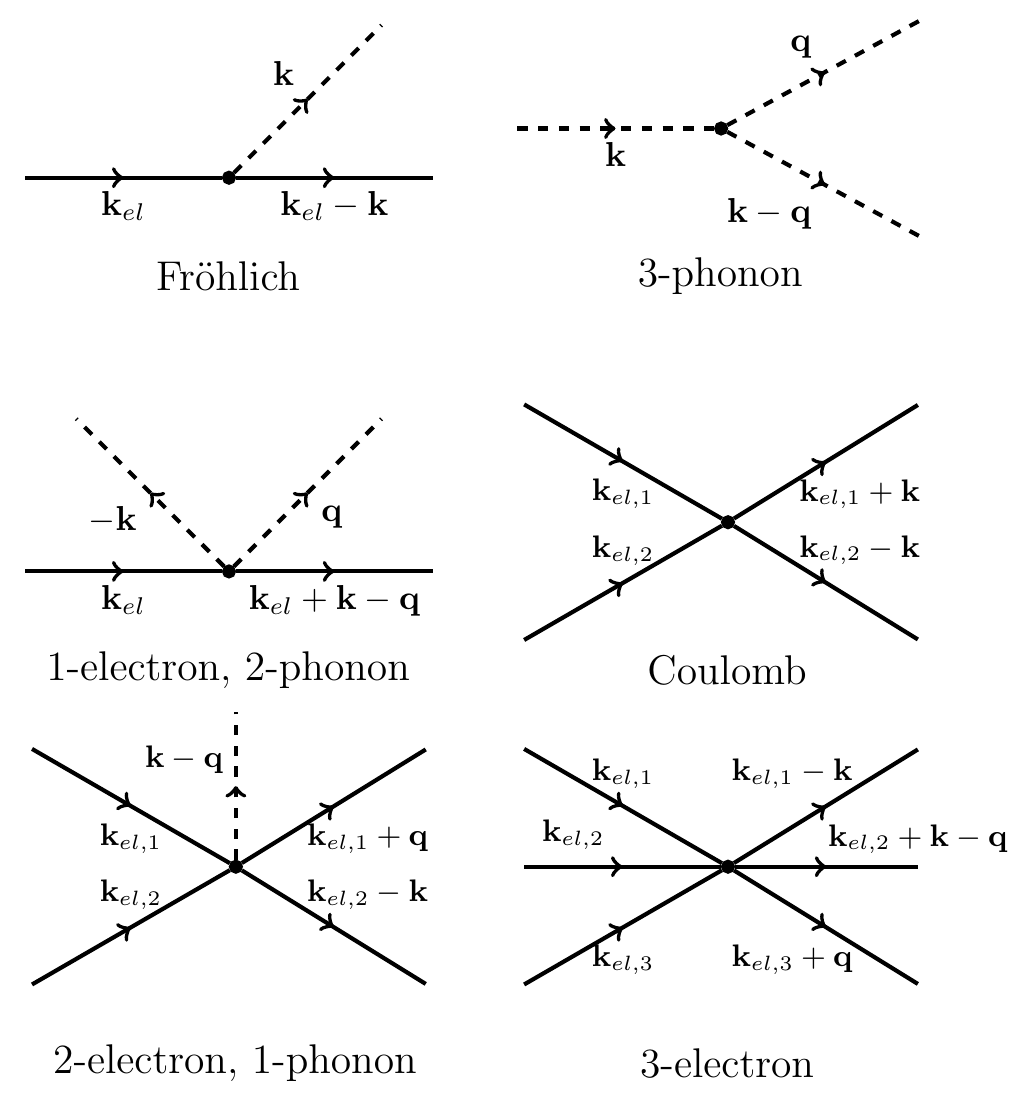}
\caption{\label{fig:HamRepresentation} A representation of the different interaction terms 
\eqref{HamGen} using Feynman diagrams. Solid lines represent electrons,
and dashed lines represent phonons. In these diagrams, any phonon line may have its arrow
and momentum reversed to create a new diagram: e.g. for the Fr\"ohlich interaction, the
electron can also absorb a phonon with momentum $-\mathbf{k}$. For a single polaron, only
the first three interactions must be considered. The 3-electron interaction is linked
to the nonlinear optical response and can therefore be neglected in most materials.}
\end{figure}

To find an expression for $\mathbf{w}$ and $\dot{\mathbf{w}}$ in terms of the ladder operators,
we introduce an auxiliary field $\mathbf{B}(\mathbf{r})$ and its Fourier transform $\op{k}$
after Fr\"ohlich \cite{Frohlich1954}:
\begin{align}
\mathbf{B}(\mathbf{r}) & = \frac{1}{i} \sqrt{\frac{1}{2\hbar}} \bm{\Omega}^{\frac{1}{2}} \left( \mathbf{w}(\mathbf{r}) + i \bm{\Omega}^{-1} \dot{\mathbf{w}}(\mathbf{r}) \right), \\
\mathbf{B}(\mathbf{r}) & = \frac{1}{\sqrt{V}} \sum_{\mathbf{k}\neq \mathbf{0}} \mathbf{n}^{\mathbf{k}} \op{k} e^{i \mathbf{k} \cdot \mathbf{r}}.
\end{align}
The equations for $\mathbf{B}(\mathbf{r})$ and $\mathbf{B}^*(\mathbf{r})$ can be inverted to obtain the following explicit expressions for $\mathbf{w}$ and $\dot{\mathbf{w}}$:
\begin{align}
\mathbf{w}(\mathbf{r}) & = -i \sqrt{\frac{\hbar}{2 V}} \bm{\Omega}^{-\frac{1}{2}} \sum_{\mathbf{k}\neq \mathbf{0}} \mathbf{n}^{\mathbf{k}} (\opc{k} + \op{-k}) e^{-i \mathbf{k} \cdot \mathbf{r}}, \label{wExpr} \\
\dot{\mathbf{w}}(\mathbf{r}) & = \sqrt{\frac{\hbar}{2 V}} \bm{\Omega}^{\frac{1}{2}} \sum_{\mathbf{k}\neq \mathbf{0}} \mathbf{n}^{\mathbf{k}} (\opc{k} - \op{-k}) e^{-i \mathbf{k} \cdot \mathbf{r}}. \label{wDotExpr}
\end{align}
These expressions will be used to eliminate $\mathbf{w}$ in the Hamiltonian \eqref{HamClassical}.
Classically, $\op{k}$ and $\opc{k}$ are the Fourier transforms of the unknown
auxiliary fields $\mathbf{B}(\mathbf{r})$ and $\mathbf{B}^*(\mathbf{r})$. In order to quantize the phonon field,
we have to impose the canonical commutation relations:
\begin{equation}
[w_j(\mathbf{r}),\dot{w}_k(\mathbf{r}')] = i\hbar \delta_{jk} \delta(\mathbf{r}-\mathbf{r}').
\end{equation}
It can be shown \cite{Frohlich1954} that these canonical commutation relations hold if
$\op{k}$ and $\opc{k}$ satisfy the bosonic commutation relations.
Therefore, we can interpret $\opc{k}$ and $\op{k}$ as the creation and annihilation
operator of the phonon field, respectively. To complete the quantisation we have to turn
$\mathbf{r}_{\text{el},i}$ and $\mathbf{p}_{\text{el},i}$ into operators, which obey the
usual commutation relations.
We now combine equations \eqref{UInternal} for the interaction energy density,
\eqref{HamClassical} for the classical Hamiltonian, \eqref{Dfourier} for the
electric displacement field, and (\ref{wExpr}-\ref{wDotExpr}) for $\mathbf{w}$ and
$\dot{\mathbf{w}}$ in order to find the quantum mechanical Hamiltonian. The volume
integrals are all of the form:
\begin{equation}
\int_V e^{i \mathbf{K}\cdot \mathbf{r}} d^3\mathbf{r} = V \delta_{\mathbf{K},0}.
\end{equation}
Then, a straightforward calculation gives the following Hamiltonian:
\begin{align}
\hat{H} & = \sum_{i=1}^{N}\frac{\hat{\mathbf{p}}^2_{\text{el},i}}{2m} + \sum_{\mathbf{k}} \hbar \omega_{\mathbf{k}} \left( \opc{k} \op{k} + \frac{1}{2} \right) +  \frac{1}{2}\sum_{\mathbf{k} \neq \mathbf{0}} V^{(C)}_{\mathbf{k}} \hat{\rho}_{\mathbf{k}} \hat{\rho}_{-\mathbf{k}} + \sum_{\mathbf{k} \neq \mathbf{0}} V^{(F)}_{\mathbf{k}} \left( \opc{k} + \op{-k} \right) \hat{\rho}_{-\mathbf{k}} \nonumber \\
& +  \sum_{\mathbf{k} \neq \mathbf{q} \neq \mathbf{0}} V^{(0)}_{\mathbf{k},\mathbf{q}} \left( \opc{-k} + \op{k} \right)\left( \opc{k-q} + \op{-k+q} \right)\left( \opc{q} + \op{-q} \right) \nonumber \\
& +  \sum_{\mathbf{k} \neq \mathbf{q} \neq \mathbf{0}} V^{(1)}_{\mathbf{k},\mathbf{q}} \left( \opc{-k} + \op{k} \right)\left( \opc{q} + \op{-q} \right) \hat{\rho}_{\mathbf{k}-\mathbf{q}} \nonumber \\
& + \sum_{\mathbf{k} \neq \mathbf{q} \neq \mathbf{0}} V^{(2)}_{\mathbf{k},\mathbf{q}} \left( \opc{k-q} + \op{-k+q} \right)\hat{\rho}_{-\mathbf{k}} \hat{\rho}_{\mathbf{q}} \nonumber \\
& + \sum_{\mathbf{k} \neq \mathbf{q} \neq \mathbf{0}} V^{(3)}_{\mathbf{k},\mathbf{q}} \hat{\rho}_{-\mathbf{k}}\hat{\rho}_{\mathbf{k}-\mathbf{q}} \hat{\rho}_{\mathbf{q}}, \label{HamGen}
\end{align}
where the sums exclude the cases where $\mathbf{k} = \mathbf{0}, \mathbf{q} = \mathbf{0}$
and $\mathbf{k} = \mathbf{q}$; this can also be taken into account by requiring that
$V^{(F)}_{\mathbf{0}} = 0$, $V^{(n)}_{\mathbf{k},\mathbf{k}} = 0$, and so on.
The phonon frequency $\omega_{\mathbf{k}}$ is given by:
\begin{equation} \label{OmegaGen}
\omega_{\mathbf{k}} = \mathbf{n}^{\mathbf{k}} \cdot \bm{\Omega} \cdot \mathbf{n}^{\mathbf{k}},
\end{equation}
Furthermore, we define $\bm{\Lambda}:=\bm{\Omega}^{-\frac{1}{2}} = \bm{\gamma}_0^{-\frac{1}{4}}$ and
the following interaction strengths, which have the dimensions of energy
and are analytical functions of $\mathbf{k}$ and $\mathbf{q}$:
\begin{align}
V^{(C)}_{\mathbf{k}} & = \frac{e^2}{V} \frac{\mathbf{n}^{\mathbf{k}} \cdot \bm{\gamma}_2 \cdot \mathbf{n}^{\mathbf{k}}}{|\mathbf{k}|^2}, \label{VCoulomb} \\
V^{(F)}_{\mathbf{k}} & = \sqrt{\frac{\hbar e^2}{2V}} \frac{\mathbf{n}^{\mathbf{k}} \cdot (\bm{\Lambda} \cdot \bm{\gamma}_1) \cdot \mathbf{n}^{\mathbf{k}}}{|\mathbf{k}|},\label{VFrohlich} \\
V^{(0)}_{\mathbf{k},\mathbf{q}} & = \frac{-i}{6\sqrt{V}} \left(\frac{\hbar}{2}\right)^{\frac{3}{2}} \left[ \Lambda_{i a} \Lambda_{j b} \Lambda_{l c} A^{(0)}_{abc} \right] n_i^{\mathbf{k}} n_j^{\mathbf{k}-\mathbf{q}} n_l^{\mathbf{q}}, \label{VAnhar0} \\
V^{(1)}_{\mathbf{k},\mathbf{q}} & = \frac{-i e \hbar}{4 V} \left[\Lambda_{i a} \Lambda_{j b} A^{(1)}_{abl} \right] \frac{n_i^{\mathbf{k}} n_j^{\mathbf{q}} n_l^{\mathbf{k}-\mathbf{q}}}{|\mathbf{k}-\mathbf{q}|}, \label{VAnhar1} \\
V^{(2)}_{\mathbf{k},\mathbf{q}} & =  \frac{-i e^2 }{2 V^{\frac{3}{2}}} \sqrt{\frac{\hbar}{2}} \left[\Lambda_{i a} A^{(2)}_{ajl} \right] \frac{n_i^{\mathbf{k}-\mathbf{q}} n_j^{\mathbf{k}} n_l^{\mathbf{q}}}{|\mathbf{k}||\mathbf{q}|}, \label{VAnhar2} \\
V^{(3)}_{\mathbf{k},\mathbf{q}} & =  \frac{-i e^3 }{6 V^2} A^{(3)}_{ijl} \frac{n_i^{\mathbf{k}} n_j^{\mathbf{k}-\mathbf{q}} n_l^{\mathbf{q}}}{|\mathbf{k}||\mathbf{k}-\mathbf{q}||\mathbf{q}|}. \label{VAnhar3}
\end{align}
The Hamiltonian given by \eqref{HamGen} is a general Hamiltonian
for $N$ large polarons interacting with a boson bath, including interaction terms up to
third order. The first line is the Fr\"ohlich Hamiltonian \eqref{FrohHam},
extended to include multiple electrons.
We note that the kinetic energy of the electrons can be replaced
by a more general energy band, for example one with an anisotropic band mass.
The other lines are the third order anharmonic correction terms to the Hamiltonian.
All of the terms can be visualized using
Feynman diagrams, which is done in Fig. \ref{fig:HamRepresentation}.

For coupling to a single LO phonon mode in a crystal with arbitrary symmetry, the
interaction strengths appearing in this Hamiltonian are given by equations
(\ref{VCoulomb}-\ref{VAnhar3}). In order, these correspond to the Coulomb interaction \eqref{VCoulomb}, the Fr\"ohlich interaction \eqref{VFrohlich}, and the anharmonic processes involving anywhere from 0 to 3 electrons and phonons \eqref{VAnhar0}-\eqref{VAnhar3}. As expected, the Coulomb interaction strength goes as
$\frac{1}{|\mathbf{k}|^2}$ and the Fr\"ohlich interaction goes as $\frac{1}{|\mathbf{k}|}$;
however, the proportionality constant can have an angular dependence if the crystal is
not cubic. The functions $V_{\mathbf{k},\mathbf{q}}^{(n)}$ are more complicated, but still
analytical.
Expression (\ref{VAnhar1}) for the interaction strength $V^{(1)}_{\mathbf{k},\mathbf{q}}$ does not agree with the result found by Kussow \cite{Kussow2009}: we will discuss this difference in section \ref{Sec:Conclusions}.

The anharmonic interaction strengths $V_{\mathbf{k},\mathbf{q}}^{(n)}$ are all purely imaginary.
This can be understood by requiring that the Hamiltonian \eqref{HamGen}
is Hermitian, which leads to the conditions:
\begin{align}
V^{(F)*}_{\mathbf{k}} & = V^{(F)}_{-\mathbf{k}}, \\
V^{(n)*}_{\mathbf{k},\mathbf{q}} & = V^{(n)}_{-\mathbf{k},-\mathbf{q}}.
\end{align}
Since the third order terms are antisymmetric in $\mathbf{k}$ and $\mathbf{q}$, the
imaginary unit is required for a Hermitian Hamiltonian.

The energy density was expanded up to third order, which means the Hamiltonian \eqref{HamGen}
is in principle unstable if the phonon displacements become too large.
As long as the phonons are approximately harmonic and the third-order terms can be
seen as correction terms, we do not expect this situation to occur. Regardless,
a stable Hamiltonian can be obtained by expanding the interaction energy density
\eqref{UInternal} up to fourth or even higher order.
Expressions \eqref{HamClassical}, \eqref{Dfourier} and \eqref{wExpr} can then be used
to obtain a Hamiltonian up to arbitrary order.

\subsection{Symmetry constraints} \label{Sec:Symmetry}
Equation \eqref{UInternal} was proposed for an arbitrary crystal. However, the symmetry
of the crystal enforces additional constraints onto the tensors $\gamma^{(n)}_{ij}$ and
$A^{(n)}_{ijl}$ in this expression. Here, we aim to find the simplest possible form of
these tensors for a crystal with cubic symmetry.

Consider a unit cell, centered at $\mathbf{r}$, with a relative displacement $\mathbf{w}$
and electric displacement field $\mathbf{D}$. In the continuum limit, $\mathbf{w}$ and
$\mathbf{D}$ can be considered constant over the entire unit cell. If a crystal symmetry
transformation $R$ is applied to both $\mathbf{w}$ and $\mathbf{D}$, the resulting unit cell
is the same as if we had simply applied $R$ to the entire system. Therefore, it must hold that:
\begin{equation} \label{UInternalReq}
U(R \cdot \mathbf{w}, R \cdot \mathbf{D}) = U(\mathbf{w},\mathbf{D}),
\end{equation}
for all crystal symmetries $R$. Note that only the rotational part of $R$ is relevant:
translations can be neglected since $\mathbf{w}$ and $\mathbf{D}$ vary little over the size
of one unit cell. Therefore, $R$ can be represented by a $3 \times 3$-matrix, and only the
point group $\mathscr{G}$ of the crystal must be considered.

Combining equations \eqref{UInternal} and \eqref{UInternalReq}, the following constraints
on the tensors $\gamma^{(n)}_{ij}$ and $A^{(n)}_{ijl}$ are obtained:
\begin{align}
& \forall R \in \mathscr{G}: \nonumber \\
\gamma^{(n)}_{ij} & = R_{ia} R_{j b} \gamma^{(n)}_{ab}, \label{gammaConstraint} \\
A^{(n)}_{ijk} & = R_{ia} R_{j b} R_{k c} A^{(n)}_{abc}. \label{AConstraint}
\end{align}
In other words, the tensors must be invariant under all lattice symmetry transformations.
It can be immediately verified that if $\gamma^{(n)}_{ij}$ or $A^{(n)}_{ijl}$ is invariant
under two different lattice transformations $R_1$ and $R_2$, it is also invariant under
$R_1 \cdot R_2$.

We apply these equations to two important cases: the case where the crystal has inversion
symmetry, and the case of cubic point groups. The inversion operator can be represented
with the matrix $R_{ij} = -\delta_{ij}$. Therefore, if the crystal has inversion symmetry,
equation \eqref{AConstraint} immediately gives $A^{(n)}_{ijk} = 0$. In this case, all of
the third order anharmonic terms in the Hamiltonian \eqref{HamGen} are
identically zero, and the Hamiltonian reduces to the Fr\"ohlich Hamiltonian. The scope of
this article is therefore limited to crystals without inversion symmetry; to investigate
anharmonicity in symmetric crystals, the internal energy density \eqref{UInternal} must be
expanded to fourth order.

To investigate the case of cubic symmetry, we start with the smallest cubic point group:
the symmetry group of a tetrahedron (denoted in Hermann-Mauguin notation as $23$). It is
generated by two elements:
\begin{align}
R_1 = \begin{pmatrix}
-1 & 0 & 0 \\
0 & -1 & 0 \\
0 & 0 & 1
\end{pmatrix},
& &
R_2 = \begin{pmatrix}
0 & 0 & 1 \\
1 & 0 & 0 \\
0 & 1 & 0
\end{pmatrix}.
\end{align}
Since the entire group can be generated by products of these elements, it suffices to
find tensors $\gamma_{ij}^{(n)}$ and $A^{(n)}_{ijl}$ that are invariant under these two
elements. For a second order tensor $\gamma_{ij}$, only the unit tensor is invariant under
both $R_1$ and $R_2$:
\begin{equation} \label{Tensor2Form}
\gamma^{(n)}_{ij} = \gamma_n \delta_{ij}.
\end{equation}
and so we obtain the familiar result for cubic crystals. For a third order tensor,
the calculation can be simplified by noting that all our third order tensors must be
fully symmetric, since this is implied by invariance under $R_2$ and either one of the
conditions \eqref{A1symmetry} or \eqref{A2symmetry}. Again, only one tensor satisfies
all of the implied constraints, and that is the absolute value of the Levi-Civita tensor
(denoted throughout this article with $\mathcal{E}$):
\begin{align}
A^{(n)}_{ijl} & = A_n \mathcal{E}_{ijl}, \label{Tensor3Form} \\
\mathcal{E}_{ijl} & = \left\{
\begin{array}{ll}
1 & \text{ if } i \neq j \neq l, \\
0 & \text{ otherwise.}
\end{array} \right.
\end{align}
This tensor is different from the one used in \cite{Kussow2009}: we postpone the comparison with
\cite{Kussow2009} until section \ref{Sec:Conclusions}.
The other four cubic symmetry groups ($m\bar{3}$, $432$, $\bar{4}3m$ and $m\bar{3}m$)
can all be obtained from the group $23$, by adding one or more generators.
Therefore, for all cubic crystals, the tensors $\gamma^{(n)}_{ij}$ and $A^{(n)}_{ijl}$
are of the form \eqref{Tensor2Form} and \eqref{Tensor3Form}. All of these tensors
can be described with a single scalar parameter, which is of great practical importance.
This scalar parameter can be identically equal to zero if the symmetry is too high:
if the crystal symmetry group is $m\bar{3}$, $432$ or $m\bar{3}m$, it holds that $A_n = 0$.
Therefore, for the remainder of this article, we will limit ourselves to crystals
whose point group is either $23$ or $\bar{4}3m$, in addition to the assumptions made
during the derivation in section \ref{Sec:Derivation}. The zincblende structure is an
important example of a crystal structure that satisfies all of these assumptions.

\subsection{Link to measurable material parameters} \label{Sec:Cubic}
In a cubic crystal, the parameters $\gamma_0, \gamma_1$ and $\gamma_2$ can be expressed
in terms of three familiar material properties: the longitudinal optical phonon frequency
$\omega_0$, the relative dielectric constant $\varepsilon$, and the square of the refractive
index $\varepsilon_{\infty}$. To find this correspondence, we start from the total energy
density up to second order, assuming equation \eqref{gammaConstraint} for a cubic crystal:
\begin{equation}
\mathcal{H} = \frac{1}{2} \Dot{\mathbf{w}} \cdot \Dot{\mathbf{w}} + \frac{1}{2} \gamma_0 \mathbf{w} \cdot \mathbf{w} + \gamma_1 \mathbf{w} \cdot \mathbf{D} + \frac{1}{2} \gamma_2 \mathbf{D} \cdot \mathbf{D}.
\end{equation}
If no electrons are present, $\mathbf{D}=\mathbf{0}$ and we immediately obtain
\begin{equation} \label{gamma0Link}
\gamma_0 = \omega_0^2.
\end{equation}
as before. To find $\gamma_1$ and $\gamma_2$, we derive the dielectric function for this
system. The electric field can be derived from equation \eqref{dU}, as well as an equation
of motion for $\mathbf{w}$:
\begin{align}
\mathbf{E} = \frac{\partial H}{\partial \mathbf{D}} & = \gamma_1 \mathbf{w} + \gamma_2 \mathbf{D}, \\
\ddot{\mathbf{w}} = -\frac{\partial H}{\partial \mathbf{w}} & = - \omega_0^2 \mathbf{w} - \gamma_1 \mathbf{D}.
\end{align}
We are now interested in the temporal Fourier transforms $\tilde{\mathbf{D}}(\omega)$ and $\tilde{\mathbf{E}}(\omega)$. In Fourier space, the equations of motion become:
\begin{align}
\tilde{\mathbf{E}}(\omega) & = \gamma_1 \tilde{\mathbf{w}}(\omega) + \gamma_2 \tilde{\mathbf{D}}(\omega), \\
-\omega^2 \tilde{\mathbf{w}}(\omega) & = - \omega_0^2 \tilde{\mathbf{w}}(\omega) - \gamma_1 \tilde{\mathbf{D}}(\omega).
\end{align}
$\tilde{\mathbf{w}}(\omega)$ can be eliminated from this equation, yielding a linear relation
between $\mathbf{D}$ and $\mathbf{E}$. The proportionality constant is the dielectric function,
which can be written as:
\begin{equation}
\varepsilon(\omega) = \frac{1}{\varepsilon_0 \gamma_2} \left( \frac{\omega^2 - \omega_0^2}{\omega^2 - \omega_0^2 + \frac{\gamma_1^2}{\gamma_2}} \right),
\end{equation}
where $\varepsilon_0$ is the vacu\"um permittivity. This dielectric function is of the polariton
type, where $\omega_0$ indeed plays the role of the longitudinal optical phonon frequency.
From the limits $\varepsilon = \varepsilon(0)$ and $\varepsilon_\infty=\varepsilon(+\infty)$, we obtain:
\begin{align}
\gamma_1 & = \omega_0 \sqrt{\frac{1}{\varepsilon_0 } \left( \frac{1}{\varepsilon_{\infty}}-\frac{1}{\varepsilon} \right)}, \label{gamma1Link} \\
\gamma_2 & = \frac{1}{\varepsilon_0 \varepsilon_{\infty}}. \label{gamma2Link}
\end{align}
Equations \eqref{gamma0Link}, \eqref{gamma1Link} and \eqref{gamma2Link} allow us to eliminate
the parameters $\gamma_0$, $\gamma_1$ and $\gamma_2$ for cubic crystals, in favour of the
experimentally available parameters $\omega_0$, $\varepsilon$ and $\varepsilon_{\infty}$.

This procedure also allows us to write the phonon frequency \eqref{OmegaGen} and interaction
strengths \eqref{VCoulomb}-\eqref{VAnhar3} in a simpler form, where the strength of each interaction
is characterized by a single scalar parameter. For example, the Fr\"ohlich interaction strength
can be written as:
\begin{equation}
V^{(F)}_{\mathbf{k}} = \hbar \omega_0 \sqrt{\frac{4\pi\alpha}{V} } \left(\frac{\hbar}{2m\omega_0} \right)^{1/4} \frac{1}{|\mathbf{k}|},
\end{equation} 
where the dimensionless Fr\"ohlich coupling constant $\alpha$ is defined as:
\begin{equation}
\alpha := \frac{1}{ 2 \hbar \omega_0} \frac{e^2}{4\pi \varepsilon_0} \sqrt{\frac{2m\omega_0}{\hbar}} \left( \frac{1}{\varepsilon_{\infty}}-\frac{1}{\varepsilon} \right).
\end{equation}
Both of these correspond to the well-known formulas for a single Fr\"ohlich polaron
\cite{Frohlich1954}. Analogously, four new dimensionless anharmonic coupling constants
can be defined. If we assume the third order tensors are of the form \eqref{Tensor3Form},
and define the dimensionless parameters $T_0$, $T_1$, $T_2$ and $T_3$ as follows:
\begin{equation}
T_n = \frac{(2\omega_0 \sqrt{\varepsilon_0})^n}{\hbar \omega_0} \left(\frac{1}{\varepsilon_{\infty}}-\frac{1}{\varepsilon} \right)^{-\frac{n}{2}} \left( \frac{\hbar m}{2\omega_0} \right)^{\frac{3}{4}} A_n,
\end{equation}
and introduce the typical polaron length scale:
\begin{equation} \label{LengthScale}
a_p := \sqrt{\frac{\hbar}{2m\omega_0}}
\end{equation}
then the phonon frequency and interaction strengths for a cubic crystal can be written in a
convenient analytic form:
\begin{align}
\omega_{\mathbf{k}} & = \omega_0 \label{omegaCubic} \\
V^{(C)}_{\mathbf{k}} & = \frac{e^2}{V \varepsilon_0 \varepsilon_{\infty}} \frac{1}{|\mathbf{k}|^2},
\label{VCoulombCubic} \\
V^{(F)}_{\mathbf{k}} & = \hbar \omega_0 \sqrt{\frac{4\pi\alpha a_p}{V} } \frac{1}{|\mathbf{k}|}, \label{VFrohlichCubic} \\
V^{(0)}_{\mathbf{k},\mathbf{q}} & = -i\frac{ \hbar \omega_0}{6} \frac{a_p^{\frac{3}{2}}}{\sqrt{V}} T_0 \ \mathcal{E}_{ijl} n_i^{\mathbf{k}} n_j^{\mathbf{k}-\mathbf{q}} n_l^{\mathbf{q}}, \label{VAnhar0Cubic} \\
V^{(1)}_{\mathbf{k},\mathbf{q}} & = -i\frac{ \hbar \omega_0}{2} \frac{\sqrt{4\pi \alpha} a_p^2 }{V} T_1 \ \mathcal{E}_{ijl} \frac{n_i^{\mathbf{k}} n_j^{\mathbf{k}-\mathbf{q}} n_l^{\mathbf{q}}}{|\mathbf{k}-\mathbf{q}|}, \label{VAnhar1Cubic} \\
V^{(2)}_{\mathbf{k},\mathbf{q}} & = -i \frac{ \hbar \omega_0}{2} \frac{4\pi \alpha a_p^{\frac{5}{2}}}{V^{\frac{3}{2}}} T_2 \ \mathcal{E}_{ijl} \frac{n_i^{\mathbf{k}} n_j^{\mathbf{k}-\mathbf{q}} n_l^{\mathbf{q}}}{|\mathbf{k}||\mathbf{q}|}, \label{VAnhar2Cubic} \\
V^{(3)}_{\mathbf{k},\mathbf{q}} & =  -i \frac{ \hbar \omega_0}{6} \frac{(4\pi \alpha)^{\frac{3}{2}} a_p^3}{V^{2}} T_3 \ \mathcal{E}_{ijl} \frac{n_i^{\mathbf{k}} n_j^{\mathbf{k}-\mathbf{q}} n_l^{\mathbf{q}}}{|\mathbf{k}||\mathbf{k}-\mathbf{q}||\mathbf{q}|}. \label{VAnhar3Cubic}
\end{align}
All of these interaction strengths consist of a prefactor that fixes the units, a dimensionless scalar
representing the relative strength of the interaction, and an analytic function of $\mathbf{k}$ and/or
$\mathbf{q}$. Unlike the coupling constant $\alpha$, the anharmonic constants $T_0$, $T_1$, $T_2$ and
$T_3$ cannot be readily written in terms of measurable quantities; however, they can still be obtained
from first principle calculations.

It must be noted that, unlike the Coulomb and Fr\"ohlich interactions, the anharmonic interaction strengths
are no longer spherically symmetric. Indeed, they have an angular dependence through the components of the
unit vectors $n_i^{\mathbf{k}}$, $n_j^{\mathbf{k}-\mathbf{q}}$ and $n_l^{\mathbf{q}}$. Despite this, the
dependence on $\mathbf{k}$ and $\mathbf{q}$ is analytic, making these interaction strengths well suited for
further theoretical investigations.

For the remainder of this article, we will consider a single polaron. In this case, the density operator
becomes:
\begin{equation}
\hat{\rho}_{\mathbf{k}} = e^{i \mathbf{k} \cdot \hat{\mathbf{r}}_{\text{el}}}.
\end{equation}
The diagrams on the bottom row of figure \ref{fig:HamRepresentation} all require more than 1 electron, since
the electron cannot interact with its own field. Therefore, the three terms in the Hamiltonian corresponding
to these diagrams drop out, and we obtain the following simplified Hamiltonian for a single polaron:
\begin{align}
\hat{H} & := \hat{H}_{\text{free}} + \hat{H}_F + \hat{H}_0 + \hat{H}_1 \label{Ham1Terms} \\
\hat{H}_{\text{free}} & := \frac{\hat{\mathbf{p}}^2_{\text{el}}}{2m} + \sum_{\mathbf{k}} \hbar \omega_{\mathbf{k}} \left( \opc{k} \op{k} + \frac{1}{2} \right) \label{Ham1Exact} \\
\hat{H}_F & := \sum_{\mathbf{k} \neq \mathbf{0}} V^{(F)}_{\mathbf{k}} \left( \opc{k} + \op{-k} \right) \hat{\rho}_{-\mathbf{k}} \label{Ham1Frohlich} \\
\hat{H}_0 & :=  \sum_{\mathbf{k} \neq \mathbf{q} \neq \mathbf{0}} V^{(0)}_{\mathbf{k},\mathbf{q}} \left( \opc{-k} + \op{k} \right)\left( \opc{k-q} + \op{-k+q} \right)\left( \opc{q} + \op{-q} \right)  \label{Ham1Anhar0}\\
\hat{H}_1 & :=  \sum_{\mathbf{k} \neq \mathbf{q} \neq \mathbf{0}} V^{(1)}_{\mathbf{k},\mathbf{q}} \left( \opc{-k} + \op{k} \right)\left( \opc{q} + \op{-q} \right) \hat{\rho}_{\mathbf{k}-\mathbf{q}}  \label{Ham1Anhar1},
\end{align}
This Hamiltonian, along with the interaction strengths \eqref{VFrohlichCubic}-\eqref{VAnhar1Cubic}, is the
central result of this article. It is the lowest order generalisation to the Fr\"ohlich Hamiltonian
\eqref{FrohHam}, making all assumptions of its derivation except for the harmonic approximation. Omitting
this approximation gives, to lowest order, two additional interaction terms. The first term,
\eqref{Ham1Anhar0}, is a 3-phonon interaction term due to the anharmonicity of the phonons. The second term,
\eqref{Ham1Anhar1}, is an ``extended'' interaction term, similar to the Fr\"ohlich interaction but involving
two phonons. Since the anharmonicity of the phonons can be included in the phonon frequency through
first-principles calculations \cite{Errea2013,Errea2014,Errea2015}, the extended interaction term is the more
interesting of the two.

\section{Perturbation theory} \label{Sec:Perturbation}
In this section, we calculate the polaron energy using perturbation theory, up to first order
in $\alpha$ and up to second order in $T_0$ and $T_1$.
The Hamiltonian \eqref{Ham1Frohlich}-\eqref{Ham1Anhar1} can be written as a sum of four contributions
\eqref{Ham1Exact}-\eqref{Ham1Anhar1}. The first contribution has known eigenstates and energy eigenvalues,
while the other three terms are interaction terms and can be considered ``small''. 

\begin{figure}
\centering
\includegraphics[width=8.6cm]{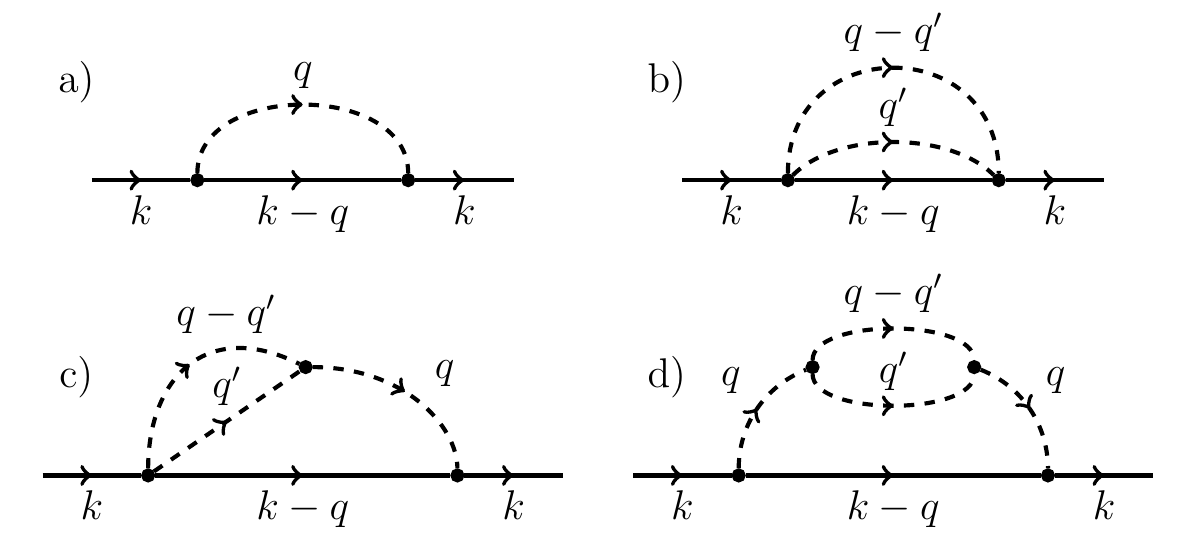}
\caption{\label{fig:PerturbationDiagrams} Diagrammatic representation of the contributions to the polaron
self energy, at zero temperature, and to first order in $\alpha$ and second order in $T_0$ and $T_1$.
The sunset diagram, shown in a),
gives the energy of the Fr\"ohlich polaron. Diagrams b) through d) contain two anharmonic interactions.
Diagram c) appears with an extra factor 2 since the order of the interactions can be switched.}
\end{figure}

The self energy contributions are shown in figure \ref{fig:PerturbationDiagrams}. Full lines
represent the electron Green's function $G_0$, dashed lines correspond to the phonon Green's
function $D_0$. Two new vertex factors are introduced: $3\sqrt{2} V^{(0)}_{\mathbf{k},\mathbf{q}}$ for the 3-phonon vertex, and $\sqrt{2} V^{(1)}_{\mathbf{k},\mathbf{q}}$ for the vertex representing
the absorption/emission of two phonons. We obtain for the self-energies of the different diagrams:
\begin{align}
\Sigma_a(\mathbf{k},\omega) & = \frac{i}{\hbar^2} \int_{-\infty}^{+\infty} \frac{d\nu}{2\pi} \sum_{\mathbf{q}}  \left|V_{\mathbf{q}}^{(F)}\right|^2 G_0(\mathbf{k}-\mathbf{q},\omega-\nu) D_0(\nu), \label{DiagAContribution} \\
\Sigma_b(\mathbf{k},\omega) & = \frac{2 i^2}{\hbar^2} \int_{-\infty}^{+\infty} \frac{d\nu d\nu'}{(2\pi)^2} \sum_{\mathbf{q},\mathbf{q}'} \left|V_{\mathbf{q}'-\mathbf{q},\mathbf{q}'}^{(1)}\right|^2 G_0(\mathbf{k}-\mathbf{q},\omega-\nu) D_0(\nu-\nu') D_0(\nu'),\label{DiagBContribution} \\
\Sigma_c(\mathbf{k},\omega) & = \frac{12 i^2}{\hbar^3} \int_{-\infty}^{+\infty} \frac{d\nu d\nu'}{(2\pi)^2} \sum_{\mathbf{q},\mathbf{q}'} V_{\mathbf{q}}^{(F)} V_{\mathbf{q},\mathbf{q}'}^{(0)} V_{\mathbf{q}'-\mathbf{q},\mathbf{q}'}^{(1)*} G_0(\mathbf{k}-\mathbf{q},\omega-\nu) D_0(\nu) D_0(\nu-\nu') D_0(\nu'),\label{DiagCContribution} \\
\Sigma_d(\mathbf{k},\omega) & = \frac{18 i^2}{\hbar^4} \int_{-\infty}^{+\infty} \frac{d\nu d\nu'}{(2\pi)^2} \sum_{\mathbf{q},\mathbf{q}'} \left|V_{\mathbf{q}}^{(F)}\right|^2 \left|V_{\mathbf{q},\mathbf{q}'}^{(0)}\right|^2 G_0(\mathbf{k}-\mathbf{q},\omega-\nu) D_0(\nu)^2 D_0(\nu-\nu') D_0(\nu').\label{DiagDContribution}
\end{align}
Adding all the contributions together and using the explicit forms of the interaction strengths \eqref{VFrohlichCubic}-\eqref{VAnhar1Cubic}, this self energy can be written rather compactly as follows:
\begin{align}
\Sigma(\mathbf{k},\omega) = & \frac{i}{\hbar^2} \int_{-\infty}^{+\infty} \frac{d\nu}{2\pi} \sum_{\mathbf{q}} \left|V_{\mathbf{q}}^{(F)}\right|^2 G_0(\mathbf{k}-\mathbf{q},\omega-\nu) \times \nonumber \\
& \ \times \left[ D_0(\nu) +  \frac{1}{2}\sum_{\mathbf{q}'} \left| \frac{6 V^{(0)}_{\mathbf{q},\mathbf{q}'}}{\hbar \omega_0} \right|^2 \left(\frac{T_1}{T_0} - \omega_0 D_0(\nu)\right)^2 i\int_{-\infty}^{+\infty} \frac{d\nu'}{2\pi}  D_0(\nu-\nu') D_0(\nu') \right].
\end{align}
In appendix \ref{Sec:AppIntegral}, we prove that
\begin{equation} \label{DivergentIntegral}
\sum_{\mathbf{q}'} \left| \frac{6 V^{(0)}_{\mathbf{q},\mathbf{q}'}}{\hbar \omega_0} \right|^2 = \frac{4 T_0^2}{15 \tilde{V_0}},
\end{equation}
where the dimensionless volume of the unit cell is defined by:
\begin{equation} \label{V0tilde}
\tilde{V}_0 = \frac{V_0}{a_p^3}.
\end{equation}
The remaining integrals are straightforward, and are overall very similar to the self energy integral for the Fr\"ohlich problem. Introducing the dimensionless variables $\tilde{\mathbf{k}} = a_p \mathbf{k}$ and $\tilde{\omega} = \omega/\omega_0$, the final result for the self energy becomes:
\begin{align}
\frac{\Sigma(\tilde{\mathbf{k}},\tilde{\omega})}{\omega_0} = & -\alpha \left(1+\frac{16 T_0}{45 \tilde{V}_0} \left(T_1 + \frac{T_0}{6}\right)\right) \frac{1}{\tilde{k}} \arctan\left(\frac{\tilde{k}}{\sqrt{1-\tilde{\omega}-i\delta}}\right), \label{Sigma1} \\
 & - \alpha \frac{2}{15\tilde{V}_0} \left(T_1 - \frac{2 T_0}{3}\right)^2 \frac{1}{\tilde{k}} \arctan\left(\frac{\tilde{k}}{\sqrt{2-\tilde{\omega}-i\delta}}\right), \label{Sigma2} \\
 & - \alpha \frac{4 T_0^2}{45 \tilde{V}_0} \frac{1}{\tilde{k}^2+1-\tilde{\omega}} \frac{1}{\sqrt{1-\tilde{\omega}-i\delta}} + O(\alpha^2). \label{Sigma3}
\end{align}
Up to first order in $\alpha$ the self energy correction to the dispersion, $E = \frac{\hbar^2 k^2}{2m} + \Sigma\left(\mathbf{k},\frac{\hbar k^2}{2m}\right)$, leads to:
\begin{small}
\begin{equation} \label{EnergyBand}
\frac{E(\tilde{\mathbf{k}})}{\hbar \omega_0} \approx \tilde{k}^2 -\alpha \left(1+\frac{16 T_0}{45 \tilde{V}_0} \left(T_1 + \frac{T_0}{6}\right)\right) \frac{ \arcsin(\tilde{k})}{\tilde{k}} + \frac{2\alpha}{15\tilde{V}_0} \left(T_1 - \frac{2 T_0}{3}\right)^2 \frac{\arcsin\left(\frac{\tilde{k}}{\sqrt{2}}\right)}{\tilde{k}} + \frac{4\alpha T_0^2}{45 \tilde{V}_0} \frac{1}{\sqrt{1-\tilde{k}^2}}.
\end{equation}
\end{small}
This energy dispersion can be expanded up to second order in $\tilde{k}$ to obtain the ground state energy and effective mass of the polaron:
\begin{align}
\frac{E_0}{\hbar \omega_0} & = -\alpha - \frac{\sqrt{2}}{15} \frac{\alpha}{\tilde{V}_0} \left[ T_1^2 + \frac{4(2\sqrt{2}-1)}{3} T_0 T_1 + \frac{2(5\sqrt{2}+2)}{9} T_0^2 \right] + O(\alpha^2), \label{E0Pert} \\
\frac{m_{\text{eff}}}{m} & = 1+\frac{\alpha}{6} + \frac{1}{90\sqrt{2}} \frac{\alpha}{\tilde{V}_0} \left[ T_1^2 + \frac{4(4\sqrt{2}-1)}{3} T_0 T_1 + \frac{4 (11\sqrt{2}+1)}{9} T_0^2 \right] + O(\alpha^2). \label{meffPert}
\end{align}
The first terms in these expressions are the well-known results for the ground state energy and effective
mass of the Fr\"ohlich polaron. The remaining terms are the corrections due to the anharmonic terms
\eqref{Ham1Anhar0}-\eqref{Ham1Anhar1}. The correction terms are proportional to $\alpha$ and
combinations of squares of the anharmonic parameters $T_0$ and $T_1$, in accordance to the results of
\cite{Kussow2009}. We find a prefactor $\tilde{V_0}^{-1}$ from the renormalisation of the integral
\eqref{DivergentIntegral}, in contrast to the prefactor $\tilde{V}_0^{-\frac{2}{3}}$ found by Kussow
\cite{Kussow2009}. This is because in \cite{Kussow2009} uses a different form for $V^{(1)}_{\textbf{k},\textbf{q}}$, as will be discussed in section \ref{Sec:Conclusions}.

\begin{figure}
\centering
\includegraphics[width=8.1cm]{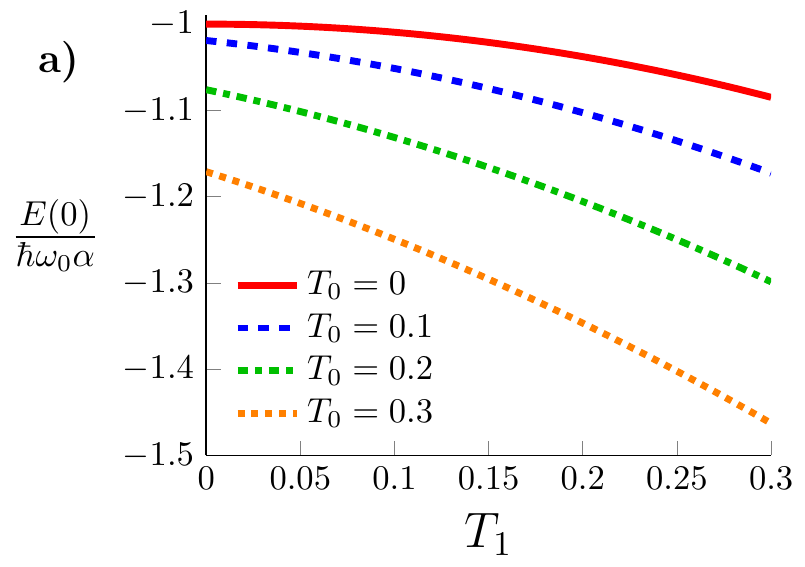}
\includegraphics[width=8.1cm]{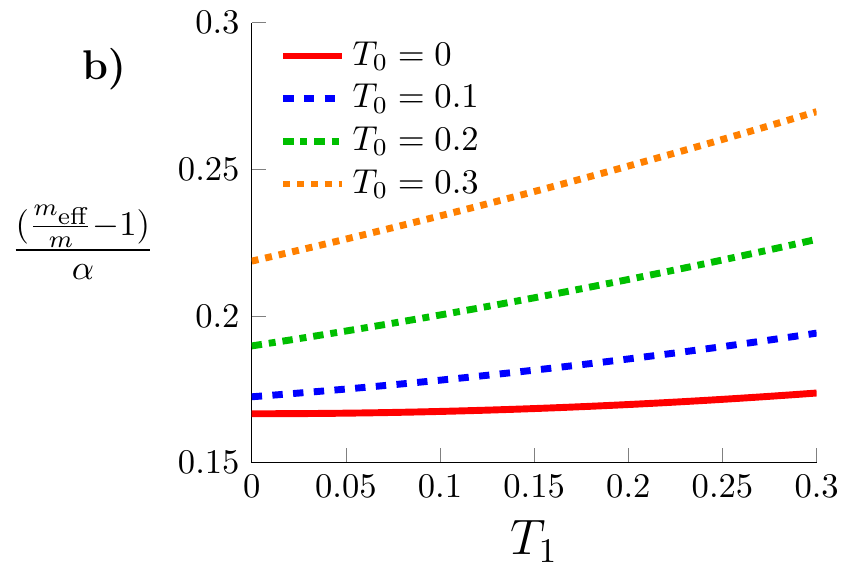}
\caption{\label{fig:PerturbationPlot} a) Ground state energy and b) effective mass of the polaron obtained
through lowest order perturbation theory, as a function of the anharmonic parameters $T_0$ and $T_1$. A
typical value of $\tilde{V}_0=0.1$ was used: this corresponds to the dimensionless volume of the zincblende
unit cell \cite{Fairbrother2014,Opoku2017,CRC}.}
\end{figure}

Figure \ref{fig:PerturbationPlot} shows the ground state energy and effective mass of the polaron in the
small coupling limit $\alpha \lesssim 1$. It is clear that the 3-phonon interaction
$V^{(0)}_{\mathbf{k},\mathbf{q}}$ and the two-phonon emmision/absorption amplitude $V^{(1)}_{\mathbf{k},\mathbf{q}}$
can both lower the ground state energy and increase the effective mass quite significantly, even for
relatively small values of $T_0$ and $T_1$.

\section{Path integral treatment} \label{Sec:Feynman}
The results from the previous section are useful in the case of weak coupling ($\alpha \lesssim 1$). However,
plenty of polar solids have stronger electron-phonon coupling. For this case, the ground state energy of the
polaron can be calculated using any of several variational methods, including the Lee-Low-Pines method
\cite{Lee1953}, the Landau-Pekar method \cite{Landau1948,Casteels2011}, and the Feynman path integral method
\cite{Feynman1955}. The Lee-Low-Pines and the Landau-Pekar methods both propose coherent phonon states as
their variational ground state, which is inadequate for the description of the extended Hamiltonian
\eqref{Ham1Frohlich}-\eqref{Ham1Anhar1}. Therefore, we will calculate the ground state energy using the path
integral method, which treats the phonons exactly and is known to give good results for the harmonic problem
at all coupling strengths \cite{Prokofev1998,Hahn2018}. Since only path integrals quadratic in the phonon
coordinates can be calculated exactly, we must limit ourselves to the case where $T_0=0$ and neglect the
three-phonon terms; however, these can be treated separately and included in a renormalized phonon frequency
\cite{Errea2013,Errea2014,Errea2015}.

\subsection{Path integral over the phonons}
The path integral method has recently been applied to the Bose polaron in ultracold gases
\cite{Ichmoukhamedov2019} where an interaction term similar to \eqref{Ham1Anhar1} is present; this derivation
follows the same general idea. This part of the derivation is valid for general functions
$\omega_{\mathbf{k}}$, $V^{(F)}_{\mathbf{k}}$ and $V^{(1)}_{\mathbf{k},\mathbf{q}}$. In the path integral
formalism, the partition sum $Z$ at finite inverse temperature $\beta$ can be written as a quantum
statistical path integral over the electron and phonon coordinates:
\begin{equation}
Z = \int \mathcal{D}\mathbf{r}(\tau) \int \mathcal{D}q_{\mathbf{k}}(\tau) \exp\left(-\frac{1}{\hbar} \int_0^{\hbar \beta} L\left(q_{\mathbf{k}}(\tau),\dot{q}_{\mathbf{k}}(\tau), \mathbf{r}(\tau),\dot{\mathbf{r}}(\tau) \right) d\tau \right),
\end{equation}
where the imaginary time Lagrangian $L$ can be found by writing equations \eqref{wExpr}-\eqref{wDotExpr} in
terms of phonon coordinates $q_{\mathbf{k}}$ and $\dot{q}_{\mathbf{k}}$ instead of creation and annihilation
operators, calculating the energy density again, and making the substitution $t \rightarrow -i \tau$.
Introducing an arbitrary phonon mass $m_{\text{ph}}$ and assuming
$V_{\mathbf{k},\mathbf{q}}^{(0)} \approx 0$, the Lagrangian is given by:
\begin{align}
 L\left(q_{\mathbf{k}}(\tau),\dot{q}_{\mathbf{k}}(\tau), \mathbf{r}(\tau),\dot{\mathbf{r}}(\tau) \right)& = \frac{m}{2} \dot{\mathbf{r}}^2 + \underset{\mathbf{k}}{\sum} \frac{m_{ph}}{2} \left(\dot{q}^*_{\mathbf{k}} \dot{q}_{\mathbf{k}} + \omega_{\mathbf{k}}^2 q^*_{\mathbf{k}} q_{\mathbf{k}} \right) + \text{Re}\left[
 \sum_{\mathbf{k}} \sqrt{\frac{2 m_{\text{ph}} \omega_{\mathbf{k}}}{\hbar}} V^{(F)}_{\mathbf{k}} \rho_{\mathbf{k}} q_{\mathbf{k}} \right] \nonumber \\
& +\text{Re}\left[ \underset{\mathbf{k},\mathbf{k}'}{\sum} \frac{2 m_{\text{ph}} \sqrt{\omega_{\mathbf{k}} \omega_{\mathbf{k}'}}}{\hbar} V^{(1)}_{\mathbf{k},\mathbf{k}'} \rho_{\mathbf{k}-\mathbf{k}'} q_{\mathbf{k}} q^*_{\mathbf{k}'}
  \right].
\end{align}
This Lagrangian is quadratic in the phonon coordinates $q_{\mathbf{k}}(\tau)$, so its path integral can be
evaluated exactly \cite{Ichmoukhamedov2019}. This is most easily done by expanding the phonon and electron
coordinates in a Fourier-Matsubara series:
\begin{align}
q_{\mathbf{k}}(\tau) & = \sum_{n} c_{\mathbf{k},n} e^{i \omega_n \tau}, \label{FourierQ} \\
\rho_{\mathbf{k}}(\tau) & = \sum_{n} f_{\mathbf{k},n} e^{i \omega_n \tau}. \label{FourierRho}
\end{align}
where the bosonic Matsubara frequencies are given by $\omega_n = \frac{2\pi n}{\hbar \beta}$.
Then, the coefficients $c_{\mathbf{k},n}$ can be integrated over the complex plane to perform the path
integral. If we consider the pair $\{\mathbf{k},n\}$ to be a single index, this integral will be a
multivariate Gaussian integral, which has a well-known expression. A straightforward calculation yields
that, if we define the ``matrix'' $A$ and the ``vector'' $B$ as follows:
\begin{align}
A_{\mathbf{k},n,\mathbf{k}',n'} & = \frac{m_{\text{ph}}}{2} (\omega_n^2 + \omega_{\mathbf{k}}^2) \delta_{\mathbf{k},\mathbf{k}'} \delta_{n,n'} + \frac{2 m_{\text{ph}} \sqrt{\omega_{\mathbf{k}} \omega_{\mathbf{k}'}}}{\hbar} V^{(1)}_{\mathbf{k},\mathbf{k}'} f_{\mathbf{k}-\mathbf{k}',n-n'}, \\
B_{\mathbf{k},n} & = \sqrt{\frac{2 m_{\text{ph}} \omega_{\mathbf{k}}}{\hbar}}  V^{(F)}_\mathbf{k} f_{\mathbf{k},n}.
\end{align}
then the path integral over the phonons can be written as follows:
\begin{align}
&\int \mathcal{D}q_{\mathbf{k}}(\tau) \exp\left(-\frac{1}{\hbar} \int_0^{\hbar \beta} L\left(q_{\mathbf{k}}(\tau),\dot{q}_{\mathbf{k}}(\tau), \mathbf{r}(\tau),\dot{\mathbf{r}}(\tau) \right) d\tau \right) \\
& \sim \int_{\mathbb{C}} \exp\left(-\beta \ \text{Re}\left[
\underset{\mathbf{k},\mathbf{k}',n,n'}{\sum} c_{\mathbf{k},n} A_{\mathbf{k},n,\mathbf{k}',n'} c^*_{\mathbf{k}',n'} +  \underset{\mathbf{k},n}{\sum} B_{\mathbf{k},n} c_{\mathbf{k},n}
\right]
\right)dc_{\mathbf{k},n},  \\
& \sim \frac{1}{\sqrt{\det(A)}} \exp\left(\frac{m_{\text{ph}} \beta}{2 \hbar^2} \sum_{\mathbf{k},n,\mathbf{k}',n'}\sqrt{\omega_{\mathbf{k}}\omega_{\mathbf{k}'}} V^{(F)*}_{\mathbf{k}} f^*_{\mathbf{k},n} A^{-1}_{\mathbf{k},n,\mathbf{k}',n'} f_{\mathbf{k}',n'} V^{(F)}_{\mathbf{k}'}\right).
\end{align}
The determinant in this expression can be rewritten in an exponential form, using
$\det(A) = e^{\text{Tr}[\ln(A)]}$. To continue, the inverse and logarithm of the matrix $A$ must be
calculated, which cannot be done in closed form. However, since $A$ is the sum of a diagonal matrix and an
additional small term, we can use the series definitions for the inverse and the logarithm to continue. The
prefactor can be obtained by comparing to the known case \cite{Kleinert2009} where
$\rho_{\mathbf{k}} = V^{(1)}_{\mathbf{k},\mathbf{k}'} = 0$. Additionally, expressions
\eqref{FourierQ}-\eqref{FourierRho} can be used to convert our expressions back to imaginary time. All of
these calculations are fairly straightforward, and the final result can be written in terms of an
effective action functional for only the electron:
\begin{align}
Z & = \left(\prod_{\mathbf{k}} \frac{1}{2\sinh(\frac{\hbar \beta \omega_{\mathbf{k}}}{2})}\right) \int \mathcal{D}\mathbf{r}(\tau) \exp\left(-\frac{1}{\hbar} S_{\text{eff}}[\mathbf{r}(\tau)] \right), \label{ZPart} \\
S_{\text{eff}}[\mathbf{r}(\tau)] & = \int_0^{\hbar \beta} \frac{1}{2} m \dot{\mathbf{r}}^2 d\tau -\hbar \sum_{n=0}^{+\infty} (-1)^n O_n[\mathbf{r}(\tau)] -\hbar \sum_{n=1}^{+\infty} \frac{(-1)^n}{n} \tilde{O}_n[\mathbf{r}(\tau)]. \label{SEff}
\end{align}
Similar to \cite{Ichmoukhamedov2019}, the effective action is written is a series form, and the terms $O_n$
and $\tilde{O}_n$ represent scattering processes. $O_n$ and $\tilde{O}_n$ are both dimensionless
functionals of $\mathbf{r}(\tau)$ of $n$-th order in
$V_{\mathbf{k},\mathbf{q}}^{(1)}$, and are given explicitly by the following expressions:
\begin{align}
O_n[\mathbf{r}(\tau)] := & \frac{1}{8} \left( \frac{2}{\hbar} \right)^{n+2} \sum_{\mathbf{k}_1,\ldots,\mathbf{k}_{n+1}} \int_0^{\hbar \beta} d \tau_1 \ldots \int_0^{\hbar \beta} d\tau_{n+2} V_{\mathbf{k}_1}^{(F)*} V^{(1)}_{\mathbf{k}_1,\mathbf{k}_2} \ldots V^{(1)}_{\mathbf{k}_{n},\mathbf{k}_{n+1}} V^{(F)}_{\mathbf{k}_{n+1}} \times \label{OnExpr} \\
& \hspace{-20pt} \times  \rho^*_{\mathbf{k}_1}(\tau_1)\rho_{\mathbf{k}_1-\mathbf{k}_2}(\tau_2) \ldots \rho_{\mathbf{k}_{n}-\mathbf{k}_{n+1}}(\tau_{n+1})\rho_{\mathbf{k}_{n+1}}(\tau_{n+2})  G_{\mathbf{k}_1}(\tau_1-\tau_2) \ldots G_{\mathbf{k}_{n+1}}(\tau_{n+1}-\tau_{n+2}), \nonumber  \\
\tilde{O}_n[\mathbf{r}(\tau)] := & \frac{1}{2} \left( \frac{2}{\hbar} \right)^n \sum_{\mathbf{k}_1, \ldots, \mathbf{k}_n} \int_0^{\hbar \beta} d\tau_1 \ldots \int_0^{\hbar \beta} d\tau_n V^{(1)}_{\mathbf{k}_1,\mathbf{k}_2} V^{(1)}_{\mathbf{k}_2,\mathbf{k}_3} \ldots V^{(1)}_{\mathbf{k}_n,\mathbf{k}_1} \times \label{OnTildeExpr} \\
& \hspace{-20pt} \times  \rho_{\mathbf{k}_1-\mathbf{k}_2}(\tau_1) \rho_{\mathbf{k}_2-\mathbf{k}_3}(\tau_2) \ldots \rho_{\mathbf{k}_n-\mathbf{k}_1}(\tau_n) G_{\mathbf{k}_1}(\tau_n - \tau_1) G_{\mathbf{k}_2}(\tau_1 - \tau_2)   \ldots G_{\mathbf{k}_n}(\tau_{n-1} - \tau_n), \nonumber
\end{align}
where we define the dimensionless phonon Green's function as:
\begin{align} \label{PhononGreensFunction}
G_{\mathbf{k}}(\tau) & := \frac{2\omega_{\mathbf{k}}}{\hbar \beta} \sum_n \frac{ e^{i\omega_n \tau}}{\omega_n^2 + \omega_{\mathbf{k}}^2} = \frac{\cosh\left(\omega_{\mathbf{k}}\left(\frac{\hbar \beta}{2}-|\tau|\right)\right)}{\sinh\left(\frac{\hbar \beta \omega_{\mathbf{k}}}{2}\right)}. & &(-\hbar \beta < \tau < \hbar \beta)
\end{align}
Some remarks must be made about the expressions \eqref{ZPart} and \eqref{SEff} for the partition sum and the
effective action. Firstly, the prefactor in \eqref{ZPart} is simply the  partition sum of the free phonon
field, which will contribute $\frac{\hbar \omega_{\mathbf{k}}}{2}$ to the ground state energy for each
phonon mode. This divergent ground state energy does not contain the coordinate $\mathbf{r}(\tau)$ and can
therefore be dropped. A similar thing can be said about $\tilde{O}_1[\mathbf{r}(\tau)]$. In
\cite{Ichmoukhamedov2019}, this term is dubbed the `vacuum polarization term'' and is denoted by
$\tilde{O}_0$ instead of $\tilde{O}_1$. From \eqref{OnTildeExpr}, one can see it also does not depend on the
electron coordinate:
\begin{equation}
\tilde{O}_1[\mathbf{r}(\tau)] = \beta \sum_{\mathbf{k}} V^{(1)}_{\mathbf{k},\mathbf{k}} \coth\left(  \frac{\hbar \beta \omega_{\mathbf{k}}}{2} \right).
\end{equation}
In our case, we have $V_{\mathbf{k},\mathbf{k}}^{(1)} = 0$, so the term is zero anyway: this means that in
expression \eqref{SEff}, we can let the second sum start at $n=2$.

\subsection{Variational principle for the free energy}
So far, no approximations have been made other than $V^{(0)}_{\mathbf{k},\mathbf{q}} = 0$: the phonons have
been treated exactly, and the problem is reduced to the single path integral  \eqref{ZPart} over the
electron coordinate. This path integral is too complicated to calculate analytically, even in the harmonic
problem. However, the Jensen-Feynman variational inequality \cite{Feynman1955,Kleinert2009} can be used to
estimate the free energy $F$ of the problem. Given any model system with action $S_0[\mathbf{r}(\tau)]$, it
holds that $F$ is bounded by:
\begin{equation} \label{JensenFeynman}
F \leq F_0 + \frac{1}{\hbar \beta} \langle S-S_0 \rangle ,
\end{equation}
where $F_0$ is the free energy of the model system, and the sharp brackets denote an expectation value with
respect to this model system. It is common \cite{Feynman1955,Tempere2009,Tempere2009Erratum,Ichmoukhamedov2019} to use a model
system where the electron is coupled to a fictitious ``phonon'' mass $M$ by a spring with spring constant
$M W^2$: the mass $M$ and the frequency $W$ are variational parameters. This model system is quadratic in
$\mathbf{r}(\tau)$, so it is possible to calculate the required expectation values. The action of this model
system can be written in terms of only the electron coordinate by tracing out the fictitious phonon
coordinate $\mathbf{Q}(\tau)$, yielding
\begin{equation} \label{S0def}
S_0 = \int_0^{\hbar \beta} \frac{m}{2} \dot{\mathbf{r}}(\tau)^2 d\tau + \frac{m W(\Omega^2-W^2)}{8} \int_0^{\hbar \beta} \int_0^{\hbar \beta} \frac{\cosh\left(  W\left(\frac{ \hbar \beta}{2}-|\tau-\tau'|\right) \right)}{\sinh\left( \frac{ W \hbar \beta}{2} \right)} |\mathbf{r}(\tau) - \mathbf{r}(\tau')|^2 d\tau d\tau',
\end{equation}
where $\Omega := W \sqrt{1+\frac{M}{m}}$ replaces $M$ as variational parameter.

All the expectation values relevant to this article can be calculated from the memory
function\cite{Tempere2009,Tempere2009Erratum,Ichmoukhamedov2019}, which is given by:
\begin{equation} \label{MemoryFunction}
\left\langle \rho^*_{\mathbf{k}}(\tau) \rho_{\mathbf{k}}(\tau') \right\rangle := \exp\left(- \frac{\hbar}{2m} k^2 D(\tau-\tau')\right),
\end{equation}
where the function $D(\tau)$ is defined as:
\begin{equation}
D(\tau) := \frac{W^2}{\Omega^2} \left(|\tau| - \frac{|\tau|^2}{\hbar \beta} \right) + \frac{1}{\Omega}\left(1-\frac{W^2}{\Omega^2} \right) \frac{\cosh\left(\Omega \frac{\hbar \beta}{2}\right)-\cosh\left(\Omega \left(\frac{\hbar \beta}{2}-|\tau|\right)\right)}{\sinh\left(\Omega \frac{\hbar \beta}{2}\right)}.
\end{equation}
From this expectation value, the free energy $F_0$ and the expectation value $\langle S_0 \rangle$ can be
evaluated exactly. These quantities only depend on the model system and not on the effective action
\eqref{SEff}, and have been calculated before \cite{Feynman1990,Tempere2009,Tempere2009Erratum,Ichmoukhamedov2019}.
Using these in the Jensen-Feynman inequality \eqref{JensenFeynman}, the variational upper bound
for the polaron free energy can be written as:
\begin{align}
F & \leq \frac{3}{\beta} \ln\left(\frac{W}{\Omega} \frac{\sinh\left(\frac{\hbar \beta \Omega}{2}\right)}{\sinh\left(\frac{\hbar \beta W}{2}\right)} \right) - \frac{3}{4} \hbar \Omega \left(1-\frac{W^2}{\Omega^2} \right)\left[ \coth\left( \frac{\hbar \beta \Omega}{2}\right)- \frac{2}{\hbar \beta \Omega}  \right] \nonumber \\
& \hspace{15pt} - \frac{1}{\beta} \sum_{n=0}^{+\infty} (-1)^n \langle O_n \rangle -\frac{1}{\beta} \sum_{n=2}^{+\infty} \frac{(-1)^n}{n} \langle \tilde{O}_n \rangle.
\end{align}
Finally, taking the temperature zero limit ($\beta \rightarrow +\infty$), the following variational
principle for the ground state energy of the polaron is obtained:
\begin{equation} \label{E0VariationalGeneral}
\frac{E(0)}{\hbar \omega_0} \leq \frac{3}{4 \omega_0} \frac{(\Omega - W)^2}{\Omega} - \lim_{\beta \rightarrow +\infty} \frac{1}{\hbar \omega_0 \beta} \sum_{n=0}^{+\infty} (-1)^n \langle O_n \rangle - \lim_{\beta \rightarrow +\infty}\frac{1}{\hbar \omega_0 \beta} \sum_{n=2}^{+\infty} \frac{(-1)^n}{n} \langle \tilde{O}_n \rangle.
\end{equation}
The problem is reduced to calculating the expectation values $\langle O_n \rangle$ and
$\langle \tilde{O}_n \rangle$ with respect to the model action using equation \eqref{MemoryFunction}. These
expectation values will be functions of the variational parameters $W$ and $\Omega$. Once these expectation
values are calculated, \eqref{E0VariationalGeneral} can be minimized with respect to $W$ and $\Omega$ to
obtain an estimate of the polaron ground state energy.

\subsection{Calculation of the expectation values}
The calculation of the general expectation values $\langle O_n \rangle$ and $\langle \tilde{O}_n \rangle$ is
a hard problem for the interaction strengths given by \eqref{VFrohlichCubic} and \eqref{VAnhar1Cubic},
mostly due to the high dimensional integrals that appear. In \cite{Ichmoukhamedov2019}, a random phase
approximation is made that allows analytical resummation of expression \eqref{E0VariationalGeneral} if the
interaction strength factorizes as
$V^{(1)}_{\mathbf{k},\mathbf{q}} \sim  V^{(F)}_{\mathbf{k}} V^{(F)}_\mathbf{q}$. Since this is not the case
for \eqref{VAnhar1Cubic}, we will instead consider the case where $T_1$ is small
and calculate only the contributions to the ground state energy op to order $T_1^2$.

First, we note that for the interaction strengths given by \eqref{VFrohlichCubic}-\eqref{VAnhar1Cubic}, the
odd order expectation values $\langle O_{2n+1} \rangle$ and $\langle \tilde{O}_{2n+1} \rangle$ are zero due
to antisymmetry. This means only $\langle O_0 \rangle$, $\langle O_2 \rangle$, and
$\langle \tilde{O}_2 \rangle$ have to be calculated. $\langle O_0 \rangle$ is the contribution from the
Fr\"ohlich action and can be calculated straightforwardly using \eqref{MemoryFunction}. Similarly,
$\langle \tilde{O}_2 \rangle$ can be calculated using the \eqref{MemoryFunction} and the result from
appendix \ref{Sec:AppIntegral}. The results are:
\begin{align}
\langle O_0 \rangle & = \hbar \omega_0 \beta \frac{\alpha}{\sqrt{\pi}} \int_0^{\frac{\hbar \beta}{2}} \frac{G(\tau)}{\sqrt{\frac{D(\tau)}{\omega_0}}} d\tau, \\
\langle \tilde{O}_2 \rangle & = \hbar \omega_0 \beta \frac{4}{15 \sqrt{\pi}} \frac{\alpha T_1^2}{\tilde{V}_0} \int_0^{\frac{\hbar \beta}{2}} \frac{G(\tau)^2}{\sqrt{\frac{D(\tau)}{\omega_0}}}  d\tau.
\end{align}
As before, the volume of the unit cell $\tilde{V}_0$ appears to renormalize the divergent integral
\eqref{DivergentIntegral}, so $\langle \tilde{O}_2 \rangle$ will be large in the continuum limit.
$\langle O_2 \rangle$ is also of order $T_1^2$ and is quite difficult to compute, but does not contain this
factor $\tilde{V}_0$ and can therefore be neglected.
With these expectation values, equation \eqref{E0VariationalGeneral} for the variational upper bound becomes:
\begin{equation} \label{E0Result}
\frac{E_0}{\hbar \omega_0} \lesssim \frac{3}{4} \frac{(v-w)^2}{v} - \frac{\alpha}{\sqrt{\pi}} \int_0^{+\infty}  \frac{e^{-\sigma} + \frac{2}{15} \frac{T_1^2}{\tilde{V}_0} e^{-2\sigma}}{\sqrt{\frac{w^2}{v^2} \sigma + \left(1-\frac{w^2}{v^2}\right) \frac{1-e^{-v\sigma}}{v}}} d\sigma.
\end{equation}
where $w := \frac{W}{\omega_0}$ and $v := \frac{\Omega}{\omega_0}$ are the new dimensionless variational
parameters. The result reduces immediately to the Feynman ground state energy \cite{Feynman1955} if
$T_1 = 0$. It must be noted that the variational inequality may no longer hold, since several terms which
may be positive were neglected in \eqref{E0VariationalGeneral}.

\begin{figure}
\centering
\includegraphics[width=8.1cm]{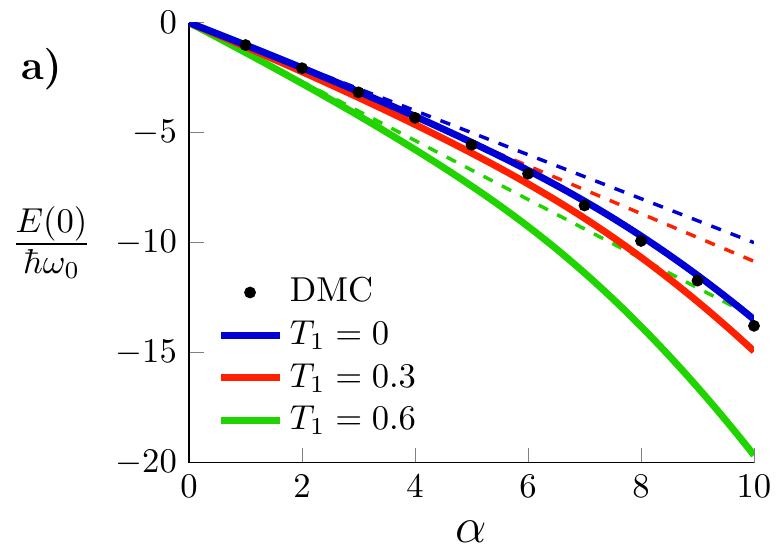}
\includegraphics[width=8.1cm]{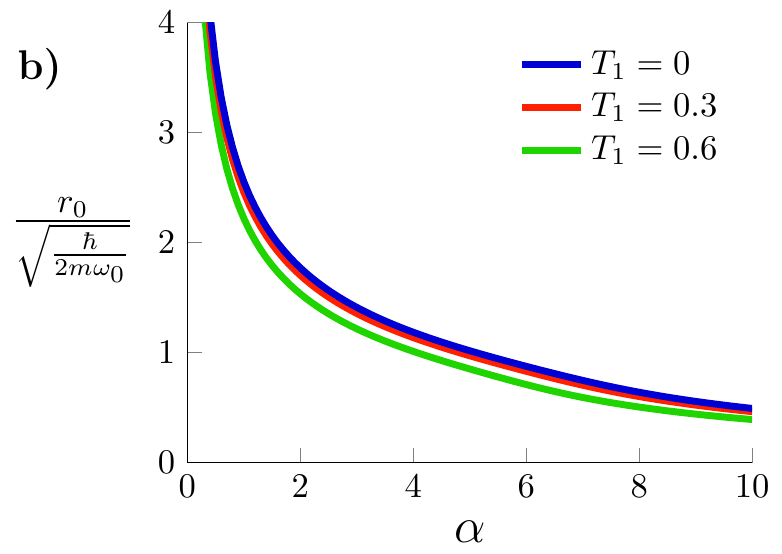}
\includegraphics[width=8.1cm]{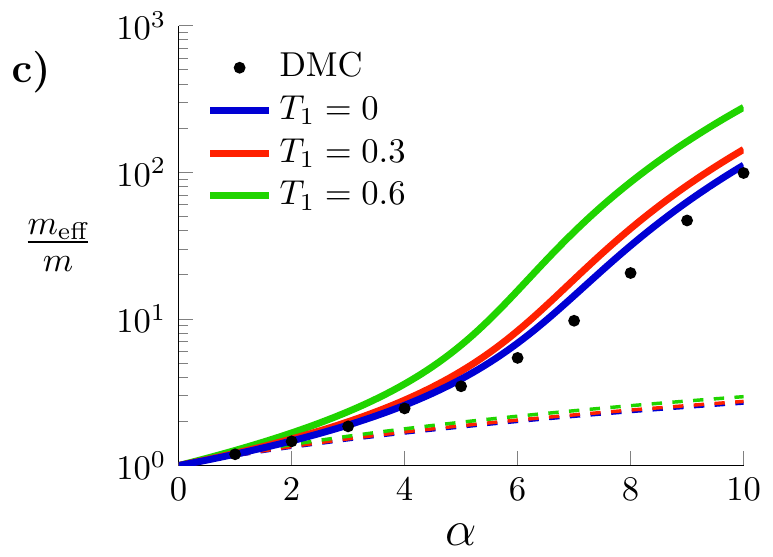}
\caption{\label{fig:Feynman} a) Ground state energy, b) radius, and c) effective mass of the polaron from
Feynman's variational method with $T_0 = 0$ and $\tilde{V}_0 = 0.1$.
Dashed lines represent the perturbation theory result \eqref{E0Pert}-\eqref{meffPert}.
Note that the effective mass is plotted on a logaritmic scale.
Figures a) and c) also shows the result obtained by the Diagrammatic Monte Carlo method \cite{Hahn2018}
for the harmonic problem ($T_1 = 0$). This shows the remarkable accuracy of the
Feynman variational method for the ground state energy at all coupling strengths, but also reveals the
inaccuracy of the effective mass in the intermediate coupling regime.}
\end{figure}

The ground state energy is obtained by numerically minimizing equation \eqref{E0Result}: the result is shown
in figure \ref{fig:Feynman}a). In the weak coupling limit $\alpha \ll 1$, the ground state energy is
minimized by $v=w$ \cite{Feynman1955}: then \eqref{E0Result} reduces to the perturbation theory
result \eqref{E0Pert} with $T_0 = 0$ as can be seen on figure \ref{fig:Feynman}a).
Similar to the perturbation theory result, the ground
state energy is significantly lowered by the anharmonic interaction. This effect is even more dramatic
in the strong coupling regime $\alpha \gg 1$. 

Once the variational parameters $v$ and $w$ are chosen in such a way that they minimize the ground state
energy, they may be used to calculated a range of other properties of the polaron.
One of these is the
polaron radius, which may be defined in several different ways.
Here, we define it using the average displacement of the relative coordinate in the model system
\cite{Schultz1959,Tempere2009,Tempere2009Erratum}:
\begin{equation}
r_0 = \sqrt{ \langle |\mathbf{r}(\tau)-\mathbf{Q}(\tau)|^2 \rangle } = \sqrt{\frac{\hbar}{2m\omega_0}} \ \frac{3 v}{v^2-w^2} \coth\left(\frac{\hbar \beta \omega_0}{2} v \right).
\end{equation}
Another is the effective mass, which can be estimated by replacing the memory function
\eqref{MemoryFunction} by the following expression, where $\mathbf{u}$ represents the velocity of the
electron \cite{Feynman1955,Tempere2009,Tempere2009Erratum,Ichmoukhamedov2019}:
\begin{equation}
\left\langle \rho^*_{\mathbf{k}}(\tau) \rho_{\mathbf{k}}(\tau') \right\rangle := \exp\left(- \frac{\hbar}{2m} k^2 D(\tau-\tau')+i \mathbf{k}\cdot\mathbf{u} \ (\tau-\tau')\right).
\end{equation}
The calculations have to be redone with this form of the memory function, up to order $u^2$. Eventually, the
ground state energy will gain an extra term of the form $\frac{1}{2}m_{\text{eff}} u^2$, where the prefactor
can be interpreted as the effective mass of the polaron. The resulting expression is:
\begin{equation}
\frac{m_{\text{eff}}}{m} = 1+ \frac{\alpha}{3\sqrt{\pi}} \int_0^{+\infty}  \frac{ \sigma^2 \left(e^{-\sigma} + \frac{2}{15} \frac{T_1^2}{\tilde{V}_0} e^{-2\sigma}\right)}{\left(\frac{w^2}{v^2} \sigma + \left(1-\frac{w^2}{v^2}\right) \frac{1-e^{-v\sigma}}{v}\right)^{\frac{3}{2}}} d\sigma.
\end{equation}

The polaron radius and the effective mass are shown in figure \ref{fig:Feynman}b) and \ref{fig:Feynman}c).
Just like the ground state energy, the effective mass is significantly increased by $T_1$ in the strong
coupling regime.
As is known from Fr\"ohlich polaron theory, the effective mass increases strongly in the intermediate
coupling regime. When the anharmonic introduction is introduced, this strong increase in the effective mass
shifts to slightly lower values of $\alpha$. 

The introduction of the anharmonic electron-phonon coupling decreases the polaron radius, as can be
intuitively expected.
The polaron radius diverges in the weak coupling limit since the polaron becomes free, and is therefore
completely delocalized. This is different from some other definitions of the polaron radius, where it is
interpreted as the spatial extent of the induced charge density \cite{Lee1953,Fedyanin1982} and therefore
remains constant as $\alpha \rightarrow 0$.

\section{Discussion and conclusions} \label{Sec:Conclusions}

The main result of this paper is the Hamiltonian \eqref{Ham1Terms}-\eqref{Ham1Anhar1} for a single
polaron in a cubic crystal with third-order anharmonicity. The derivation is presented in such a way that it
can be straightforwardly generalized to the case of multiple polarons, general crystal symmetries, and
higher order anharmonic terms. If cubic symmetry is assumed, the harmonic interaction strengths and the
phonon frequency are isotropic and reduce to the Fr\"ohlich form as expected \cite{Frohlich1954}. However,
it is not enough to make the anharmonic interaction strengths \eqref{VAnhar0Cubic}-\eqref{VAnhar3Cubic}
isotropic, and tensor notation or index notation is still required despite the cubic symmetry.

Hamiltonians of the form \eqref{Ham1Terms}-\eqref{Ham1Anhar1}, including the anharmonic interactions, can be found in other areas of physics. A notable example is the Bose polaron Hamiltonian, describing impurities in ultracold Bose gases under the Bogolioubov approximation \cite{Shchadilova2016,Ichmoukhamedov2019}. The Hamiltonians only differ in their different expressions of the interaction strengths $V^{(F)}_\mathbf{k}$, $V^{(0)}_{\mathbf{k},\mathbf{q}}$, and $V^{(1)}_{\mathbf{k},\mathbf{q}}$.

To obtain the Hamiltonian \eqref{Ham1Terms}-\eqref{Ham1Anhar1}, we have made several drastic assumptions: the material must belong to either of the point groups $23$ or $\bar{4}3m$, and the primitive unit cell must contain only two atoms. Regardless, III-V semiconductors such as BN, BP, AlN and AlP all exist in the zincblende structure and therefore satisfy all of the above assumptions. In addition, these semiconductors display significant anharmonicity \cite{Shulumba2016,Yaddanapudi2018,Brito2019} since their ions are relatively light. In any of these materials, anharmonic polarons corresponding to the Hamiltonian \eqref{Ham1Terms}-\eqref{Ham1Anhar1} may be experimentally observed, for example by measuring the mid-infrared optical conductivity. The Fr\"ohlich electron-phonon coupling \eqref{Ham1Frohlich} gives rise to an absorption peak around $\omega \approx \omega_0$ \cite{Finkenrath1969, Tempere2001, VanMechelen2008}. Since the anharmonic electron-phonon coupling \eqref{Ham1Anhar1} involves the simultaneous creation of two phonons (see figure \ref{fig:PerturbationDiagrams}b), it will give rise to a secondary peak around $\omega \approx 2 \omega_0$. This secondary peak serves as a fingerprint for anharmonicity of the form \eqref{Ham1Anhar0}-\eqref{Ham1Anhar1}.

The Hamiltonian \eqref{Ham1Terms}-\eqref{Ham1Anhar1} for a single polaron contains two unknown
dimensionless material parameters $T_0$ and $T_1$, which characterize the relative strength of the two
anharmonic interactions. Although $T_0$ can in theory be linked to the Gr\"uneisen constant $\gamma_{LO}$ of
the longitudinal optical phonons, we found no way to directly link $T_0$ and $T_1$ to experimentally
available material parameters. Therefore even an order of magnitude estimate of $T_0$ and $T_1$, and any
quantitative comparison to experiments, is difficult. In addition to this, all treatments of anharmonic
polarons besides that of Kussow \cite{Kussow2009} have focused on small polarons
\cite{Zolotaryuk1998,Voulgarakis2000,Velarde2010}, so any comparison with these results would have to be
qualitative anyway.

We based our work on Kussow \cite{Kussow2009}, who also derives a Hamiltonian of the form \eqref{Ham1Anhar1}.
Our expression for the interaction strength \eqref{VAnhar1Cubic} does not match the expression in
\cite{Kussow2009}, for two reasons. Firstly, \cite{Kussow2009} uses the following form for the tensor
$A^{(1)}_{ijk}$:
\begin{align}
A^{(1)}_{ijl} = A_1 \delta_{ijl}, &&
\text{ where } \delta_{ijl} = \left\{
\begin{array}{ll}
1 & \text{ if } i = j = l, \\
0 & \text{ otherwise}.
\end{array} \right.
\end{align}
instead of the correct form \eqref{Tensor3Form}. Additionally, the final integral in their derivation
(equation 40 in \cite{Kussow2009}) is incorrectly calculated using Green's theorem. Despite this, the
Hamiltonian in \cite{Kussow2009} is still of the correct order of magnitude, so their results should still
be qualitatively correct. Indeed, in \cite{Kussow2009} it is also concluded that the ground state energy is
lowered and the effective mass is increased by the anharmonic interaction, and that the transition to a
small polaron occurs at lower values of $\alpha$ if the anharmonic interaction is present. These qualitative
results are also seen for the small polaron if an asymmetric on-site potential is used
\cite{Zolotaryuk1998}. Interestingly, the effects of anharmonicity seem to be reversed if a quartic
potential is used \cite{Voulgarakis2000} instead of a cubic potential. It could be interesting to see if
similar results are found if the internal energy \eqref{UInternal} is expanded up to fourth order and a
crystal with inversion symmetry is considered.

The derived Hamiltonian \eqref{HamGen} is well suited for the calculation of further
anharmonic large polaron properties, since the interaction strengths are analytical and depend on only one
dimensionless parameter each. In addition, it is a direct generalization of the Fr\"ohlich Hamiltonian,
meaning most theoretical techniques for solving the Fr\"ohlich Hamiltonian can be used for this Hamiltonian
as well (except perhaps the Lee-Low-Pines and Landau-Pekar methods, as motivated in section
\ref{Sec:Feynman}). For example, the electron mobility/AC conductivity/optical response can be calculated
semi-analytically using several different methods \cite{Feynman1962,Peeters1983,Tempere2001}. The optical response is of great significance
for high-pressure hydride and metallic hydrogen experiments
\cite{Drozdov2015,Dias2017,Zaghoo2017,Loubeyre2019,Somayazulu2019} and can even be used to determine the
superconducting transition \cite{Carbotte2018}. Using this Hamiltonian it is also possible to investigate
bipolaron formation \cite{Verbist1991}, which has been proposed as a possible pairing mechanism for
superconductivity. While bipolarons can only occur at $\alpha > 6.8$ in the harmonic approximation
\cite{Verbist1991}, the increased electron-phonon interaction energy suggests a wider stability regime in
the anharmonic case. Further calculations may indicate whether bipolarons can occur for values of $\alpha$,
$T_0$ and $T_1$ corresponding to a realistic material.

\acknowledgments

This research was funded by the University Research Fund (BOF) of the University of Antwerp. We would like
to thank T. Ichmoukhamedov, T. Hahn and S. Ragni for many interesting discussions. We also thank C. Franchini and
G. Kresse from the University of Vienna for their long-standing collaboration with our research group,
and for agreeing to calculate the anharmonic coefficients from first principles. 

\appendix

\section{Integral over the anharmonic interaction strength} \label{Sec:AppIntegral}
During the calculation of the polaron energy using both perturbation theory and the path integral formalism,
we encountered the following sum:
\begin{equation}
\mathcal{I} = \sum_{\mathbf{k}} \left| \frac{6 V^{(0)}_{\mathbf{q},\mathbf{k}}}{\hbar \omega_0} \right|^2.
\end{equation}
Unlike the other integrals in this article, the calculation of the above integral is
not quite straightforward and deserves some further explanation.

The sum can be transformed into an integral using $\sum_{\mathbf{k}} \rightarrow \frac{V}{(2\pi)^3} \int d^3\mathbf{k}$. Then, we can plug in expression \eqref{VAnhar0Cubic} for $V^{(0)}_{\mathbf{q},\mathbf{k}}$ to obtain:
\begin{equation}
\mathcal{I} = \frac{T_0^2 a_p^3}{(2\pi)^3} \int \left(\mathcal{E}_{ijl} n_i^{\mathbf{k}} n_j^{\mathbf{k}-\mathbf{q}} n_l^{\mathbf{q}}\right)^2 d^3 \mathbf{k},
\end{equation}
where the indices $i,j,l$ are summed over the values $\{x,y,z\}$. Firstly, we note that the integration domain is very large and the vector $\mathbf{q}$ is finite. This
allows us to replace $n_j^{\mathbf{k}-\mathbf{q}}$ by $n_j^{\mathbf{k}}$:
\begin{equation}
\mathcal{I} = \frac{T_0^2 a_p^3}{(2\pi)^3} \int \left(\mathcal{E}_{ijl} n_i^{\mathbf{k}} n_j^{\mathbf{k}} n_l^{\mathbf{q}}\right)^2 d^3 \mathbf{k}.
\end{equation}
Formally, this can be justified by substituting $\mathbf{k} \rightarrow a \mathbf{K}$ for some large
value $a$: this substitution makes $\mathbf{q}$ negligibly small compared to $a \mathbf{K}$, but
otherwise leaves the integral invariant since $n_i^{a\mathbf{K}} = n_i^{\mathbf{K}}$. More intuitively, the
following argument can be used: since the integrand remains finite as
$|\mathbf{k}| \rightarrow +\infty$, the integral will be dominated by the domain in which
$|\mathbf{k}| \gg |\mathbf{q}|$, and in this domain the vector
$\mathbf{k}-\mathbf{q}$ and the vector $\mathbf{k}$ approximately have the same
direction.

The volume element $d^3\mathbf{k}$ can now be written in spherical coordinates, and the radial and angular
integrals can be split:
\begin{align}
\mathcal{I} & = \frac{T_0^2 a_p^3}{(2\pi)^3} \mathcal{E}_{ijl} \mathcal{E}_{abc} n_l^{\mathbf{q}} n_c^{\mathbf{q}} \int n_i^{\mathbf{k}} n_j^{\mathbf{k}}  n_a^{\mathbf{k}} n_b^{\mathbf{k}} d^3 \mathbf{k}, \\
 & = \frac{T_0^2 a_p^3}{(2\pi)^3} \mathcal{E}_{ijl} \mathcal{E}_{abc} n_l^{\mathbf{q}} n_c^{\mathbf{q}} \left(\int k^2 dk \right) \left( \int_0^{\pi} \int_0^{2\pi} n_i n_j n_a n_b \sin(\theta) d\theta d\varphi \right), \label{Int1App}
\end{align}
where $\mathbf{n} = \mathbf{n}(\theta,\varphi)$ is a unit vector in the direction defined by the angles $\theta$ and
$\varphi$. Its components are given by:
\begin{align}
n_x(\theta,\varphi) & := \sin(\theta) \cos(\varphi), \\
n_y(\theta,\varphi) & := \sin(\theta) \sin(\varphi), \\
n_z(\theta,\varphi) & := \cos(\theta).
\end{align}
The angular integral is a special case of the following integral identity \cite{Thorne1980}:
\begin{equation}
\int_0^{\pi} \int_0^{2\pi} n_{i_1} n_{i_2} \ldots n_{i_\ell} \sin(\theta) d\theta d\varphi = \left\{ \begin{array}{ll}
\frac{4\pi}{\ell + 1} \delta_{(i_1 i_2} \delta_{i_3 i_4} \ldots \delta_{i_{\ell-1} i_{\ell})} & (\ell \text{ even}) \\
0 & (\ell \text{ odd})
\end{array}
 \right. .
\end{equation}
Then expression \eqref{Int1App} becomes:
\begin{equation}
\mathcal{I} = \frac{T_0^2 a_p^3}{5(2\pi)^3} \delta_{(ij} \delta_{ab)} \mathcal{E}_{ijl} \mathcal{E}_{abc} n_l^{\mathbf{q}} n_c^{\mathbf{q}} \left(4\pi \int k^2 dk \right).
\end{equation}
The contractions over the indices $i,j,a$ and $b$ can now be performed, using the fact that
$\delta_{(ij} \delta_{ab)} \mathcal{E}_{ijl} \mathcal{E}_{abc} = \frac{4}{3} \delta_{lc}$.
Then the integral becomes:
\begin{align}
\mathcal{I} & = \frac{4 T_0^2 a_p^3}{15(2\pi)^3} \mathbf{n}^{\mathbf{q}} \cdot \mathbf{n}^{\mathbf{q}} \int k^2 dk, \\
\mathcal{I} & = \frac{4T_0^2}{15} \times \frac{a_p^3}{(2\pi)^3} \left[4\pi \int k^2 dk \right].
\end{align}
If we assume $k \in [0,+\infty]$, the radial integral obviously diverges. However, physically, we only
expect wavevectors in the first Brillouin zone to be relevant. Since we work in the continuum limit, the
first Brillouin zone will be large, but not infinite, and we can use the volume of the first Brillouin zone
as a Debye cutoff for the radial integral over $k$. The quantity between square brackets should
equal the volume of the first Brillouin zone, which is equal to
$\frac{(2\pi)^3}{V_0}$ where $V_0$ is the volume of the unit cell. Therefore, we
finally find:
\begin{equation}
\mathcal{I} = \sum_{\mathbf{k}} \left| \frac{6 V^{(0)}_{\mathbf{q},\mathbf{k}}}{\hbar \omega_0} \right|^2 = \frac{4 T_0^2}{15 \tilde{V}_0}.
\end{equation}
where we defined $\tilde{V}_0 := V_0/a_p^3$ as in equation \eqref{V0tilde}. This is the result we presented in the main text.

\bibliography{References}

\end{document}